\documentclass[reprint,
amsmath,amssymb,aps,
pre,showkeys,
floatfix]{revtex4-2}
\usepackage{dcolumn,bm,chemformula,graphicx,titlesec}
\usepackage[unicode]{hyperref}
\hypersetup{
	colorlinks= true,
	citecolor= blue,
	linkcolor= red,
	urlcolor=magenta,
	filecolor=cyan,
	linkbordercolor={1 0 0},
	citebordercolor={0 1 0},
	urlbordercolor={0 1 1},
} 
\usepackage[mathlines]{lineno}
\usepackage[section]{placeins}
\usepackage{float,wrapfig}
\usepackage[tight,TABTOPCAP]{subfigure}
\begin{document}
	\preprint{APS/123-QED}
	\title{Nonequilibrium thermodynamics of glycolytic traveling wave: Benjamin-Feir instability}
	\author{Premashis Kumar}
	\affiliation{S. N. Bose National Centre For Basic Sciences, Block-JD, Sector-III, Salt Lake, Kolkata 700 106, India}
	\author{Gautam Gangopadhyay}
	\email{gautam@bose.res.in}
	\affiliation{S. N. Bose National Centre For Basic Sciences, Block-JD, Sector-III, Salt Lake, Kolkata 700 106, India}
	\date{\today}
	\begin{abstract}

Evolution of the nonequilibrium thermodynamic entities corresponding to dynamics of the Hopf instabilities and traveling waves at a nonequilibrium steady state of a spatially extended glycolysis model is assessed here by implementing an analytically tractable scheme incorporating complex Ginzberg-Landau equation(CGLE). In the presence of self and cross diffusion, a more general amplitude equation exploiting the multiscale Krylov-Bogoliubov averaging method serves as an essential tool to reveal the various dynamical instability criteria, specially Benjamin-Feir(BF) instability, to estimate the corresponding nonlinear dispersion relation of the traveling wave pattern. The critical control parameter, wavenumber selection criteria, and magnitude of the complex amplitude for traveling waves are modified by self- and cross-diffusion coefficients within the oscillatory regime, and their variabilities are exhibited against the amplitude equation. Unlike the traveling waves, a low amplitude broad region appears for the Hopf instability in the concentration dynamics as the system phase passes through minima during its variation with the control parameter. The total entropy production rate of the uniform Hopf oscillation and glycolysis wave not only qualitatively reflects the global dynamics of concentrations of intermediate species but almost quantitatively. Despite the crucial role of diffusion in generating and shaping the traveling waves, the diffusive part of the entropy production rate has a negligible contribution to the system's total entropy production rate. The Hopf instability shows a more complex and colossal change in the energy profile of the open nonlinear system than so in the traveling waves. A detailed analysis of  BF instability shows a  contrary nature of the semigrand Gibbs free energy with discrete and continuous wavenumbers for the traveling wave. We hope the Hopf and traveling wave pattern around the BF instability in terms of energetics and dissipation would open up new applications of such dynamical phenomena.
	\end{abstract}
	
	\maketitle
\section{\label{intro}Introduction}
Oscillation is a ubiquitous phenomenon in a living system\citep{Murray2003MathematicalApplications, epstein1998introduction, Cross2009PatternSystems} starting from cellular rhythms\citep{goldbeter1997biochemical, falcke2004reading, thurley2012fundamental}, oscillation in single enzyme system\citep{olsen} to glycolytic oscillation in a cell\citep{GHOSH1964174, Boiteux3829} with a diverse oscillatory pattern from different origin but with a universal underlying principle. Glycolysis, a crucial energy-generating pathway of the metabolism in a living system, involves a complex chemical reaction network. From simple nonlinear models\citep{selkov, Goldbeter3255} to sophisticated complex models\citep{HYNNE2001121, wolf2000effect, Madsen}, several theoretical schemes have been proposed to capture the temporal and spatial oscillatory behavior of glycolysis. However, the intricacy of the sophisticated glycolysis model hinders the investigation of any particular mechanism within a specific regime.  Therefore, in the spirit of simplicity and clarity, we have chosen the two-variable Selkov model extended by diffusion, which provides scope for studying the vital dynamical features and elaborating them from the theoretical ground\citep{straube2008, Lavrova2009PhaseInflux}. 
From the kinetic picture of glycolysis to the Selkov equation, obscurity arises with the consideration of many irreversible states. Here we have concentrated on the reversible kinetics of the Selkov model as an open system with Rayleigh oscillator form to study nonequilibrium steady state(NESS)\citep{qian2006open} phenomena. 

The generation of traveling waves from glycolytic activity in the diffusive layer exploiting the yeast extracts in an open spatial reactor is previously  demonstrated\citep{MAIR1996627, BAGYAN200567} experimentally. On the other hand, an amplitude equation\citep{aranson2002world, Cross2009PatternSystems} in the presence of diffusive coupling is utilized to explain the appearance of inward rotating spiral waves in glycolysis\citep{STRAUBE2010L4, StraubeN10}. However, in the previous study\citep{Lavrova2009PhaseInflux} of the Selkov model involving amplitude equation, either diffusion coefficients are taken equally in magnitude for analytic investigation or their contribution is neglected.  In our amplitude equation consideration, all the diffusion coefficients(self and cross) are present and can take any range of values. In this aspect, our findings related to the amplitude equation would enable one with more flexibility in exploring and studying various standard glycolysis models' performance, generic features, and robustness. For the Selkov glycolysis model, inhomogeneous control parameter flux \citep{Lavrova2009PhaseInflux, LAVROVA2009127} or periodic substrate influx\citep{Verveykochaos} has been often considered to investigate various rich features like phase reversal, chaotic oscillation, or oscillation entrainment within the oscillatory regime. However, we have considered the homogeneous concentration of the control parameter in this report. Therefore, features that appear in homogenous Hopf oscillation or traveling waves are solely due to uniform chemostatted species concentration and diffusion coefficients. 

The main focus of investigating glycolysis waves was limited to controlling the spatiotemporal pattern\citep{epstein1998introduction, Cross2009PatternSystems}, recognizing the vital influential factors of the oscillating behavior and entrainment of intrinsic glycolytic oscillations till now.  However, the role of glycolysis waves in processing and spreading biological information\citep{PETTY2006217, BEHAR2010684,  PURVIS2013945}, and thus dictating the coordination among events in the system, seeks to address the questions related to entropic cost, energetics, or efficiency of the wave. In this report, we have investigated the evolution of nonequilibrium thermodynamic entities corresponding to the uniform oscillation and traveling waves in a simple glycolysis model system of finite size. This thermodynamic description will help to understand any nonlinear system containing limit cycles or waves on a fundamental level. Furthermore, our general analysis of canonical CGLE equation for the extended reaction-diffusion systems in the presence of both self and cross diffusion\citep{vanagcross}  in this report will shed new light on how the different instabilities dictated by the coefficients of amplitude equation to imprint their signatures on the evolution of the thermodynamic entities near Hopf instability point at NESS. 

The layout of the report is as follows. In sec. \ref{sec:1}, we have discussed the reaction dynamics of the Selkov model and its reversible equivalent version. The reaction-diffusion form of the glycolytic model and its linear stability analysis are provided in the next section. In sec. \ref{selampli}, we have derived the CGLE equation using the Krylov-Bogolyubov(KB) method and then separate the magnitude and the phase dynamics. In sec. \ref{cds}, the concentration fields are obtained by combining analytical results in previous sections. Entropy production rate and nonequilibrium Gibbs free energy are formulated for the reaction-diffusion system in sec. \ref{sec:thermodynamics}. We have provided numerical results and discussions in sec. \ref{redi}. Finally, the paper is concluded in sec. \ref{conclu}.

\section{\label{sec:1}Simple Glycolysis Model: From Kinetic Selkov Model to Reversible Model}
In this section, we would derive simplified partial differential equation forms of the simple glycolytic models from their standard chemical reaction networks. The simplified system will be utilized for linear stability analysis and amplitude equation formulation in later sections of the report. 
\subsection{Kinetic Selkov Model}
E. E. Selkov proposed a simple kinetic model of glycolysis\citep{selkov} that exhibits periodic oscillation for a specific range of parameters. The Selkov model contains the following sequence of chemical reactions: 
\begin{equation}
\begin{aligned}
\rho&=1:&\ch{S_1 + ES_{2}^2&<=>[\text{k_1}][\text{k_{-1}}] S_1ES_{2}^2} \\
\rho&=2:&\ch{S_1ES_{2}^2&->[\text{k_{2}}] ES_{2}^2 + S_2}\\
\rho&=3:& \ch{2 S_2 + E&<=>[\text{k_{3}}][\text{k_{-3}}]
	ES_{2}^2}\\ 
\end{aligned}
\label{crn}
\end{equation}
where $'\rho'$ is reaction step label. The substrate $S_1(ATP)$ is supplied at constant rate $z_1$ and product $S_2(ADP)$ is removed at a rate $v_2$. The free enzyme  $E(phosphofructokinase)$ is initially inactive and becomes active only  after combining with the product $S_2$ to form a complex $ES_{2}^2$. It should be noted that chemical reaction corresponding to $\rho=2$  here generates the product, $S_2(ADP)$ irresversibly. 

With the assumption, all the reverse  rate constants $k_{-\rho}$ are vanishingly small($10^{-4}$), and  the  forward  reaction  rate  constants $k_{\rho}$ are much higher than the reverse one, i.e., $k_{\rho}\gg k_{-\rho}$, rate equations of concentrations of intermediate species in eq. \eqref{crn} yield the 
\begin{equation}
\begin{aligned}
\dot{x_1}&=(k_1+k_2)x_2-k_1s_1x_1+{k_3}(e_0-x_1-x_2)s_2^2-k_{-3}x_1 \\
\dot{x_2}&={k_1}s_1x_1-(k_{-1}+k_2)x_2 
\label{dynamic}
\end{aligned}
\end{equation}
where concentrations of the species are denoted by  $$x_1=[ES_{2}^2],x_2=[S_1ES_{2}^2],s_1=[S_1],s_2=[S_2] .$$

The steady state solution of the eq. \eqref{dynamic} is 
\begin{subequations}
	\begin{align}
		x_1^{s.s.}&=\frac{e_0k_3s_2^2[k_2+k_{-1}]}{S}\label{steadyx1}\\
		x_2^{s.s.}&=\frac{e_0k_3k_1s_1s_2^2}{S}\label{steadyx2}
	\end{align}	
\end{subequations} 
where $S=(k_1s_1+k_3s_2^{2}+k_{-3})(k_2+k_{-1})-(k_{-1}+k_2-k_3s_2)(k_1s_1)$. After rearranging eq. \eqref{steadyx1} and \eqref{steadyx2} we arrive at the following form:
\begin{subequations}
	\begin{align}
		x_1^{s.s.}&=\frac{e_0\zeta_2^2}{1+\zeta_2^2(1+\zeta_1)}\label{newsteadyx1}\\
		x_2^{s.s.}&=\frac{e_0\zeta_1\zeta_2^2}{1+\zeta_2^2(1+\zeta_1)}\label{newsteadyx2}
	\end{align}	
\end{subequations}
where $\zeta_1=\frac{k_1}{k_{-1}+k_2}s_1$ and $\zeta_2=\sqrt{\frac{k_3}{k_{-3}}}s_2$ are relative concentrations of substrate  and product, respectively. 
Furthermore, we can also obtain, $e^{s. s.}=\frac{e_0}{1+\zeta_2^2(1+\zeta_1)}$ by exploiting the fact that total enzyme concentration remains constant over the whole process. From eq. \eqref{crn}, we can write following dynamical equation 
\begin{subequations}
	\begin{align}
		\frac{\partial s_1}{\partial t}&=z_	1-k_1s_1x_1+k_{-1}x_2\label{s1anotherdynamic}\\
		\frac{\partial s_2}{\partial t}	&=k_2x_2-{k_3}(e_0-x_1-x_2)s_2^2+k_{-3}x_1-k_2s_2.
		\label{anotherdynamic}		
	\end{align}
\end{subequations}	
After dimensionless analysis of eq. \eqref{s1anotherdynamic} and \eqref{anotherdynamic}, we have 
\begin{subequations}
	\begin{align}
		\frac{\partial \zeta_1}{\partial \theta}&=z-(1+\frac{k_{-1}}{k_2})\frac{\zeta_1x_1}{e_0}+\frac{k_{-1}}{k_2}\frac{x_2}{e_0}\label{zeta1anotherdynamic}\\
		\frac{\partial \zeta_2}{\partial \theta}	&=\alpha_2[\frac{x_2}{e_0}-\frac{k_{-3}}{k_2}\frac{e}{e_0}\zeta_2^2+\frac{k_{-3}}{k_2e_0}x_1-X_2\zeta_2], 
		\label{zeta2anotherdynamic}		
	\end{align}
\end{subequations}		
where $z=\frac{z_1}{k_2e_0}, \theta=\frac{k_1k_2e_0t}{k_{-1}+k_2}, \alpha_2=\frac{k_2+k_{-1}}{k_1}\sqrt{\frac{k_3}{k_{-3}}},X_2=\frac{1}{e_0}\sqrt{\frac{k_{-3}}{k_3}}$. Now further applying dimensionless analysis on eq. \eqref{dynamic}, we get 
\begin{subequations}
	\begin{align}
		\epsilon\frac{\partial x_1}{\partial \theta}&=x_2-x_1\zeta_1-\frac{K_3}{K_1+1}[x_1-\zeta_2^2e]\label{x1anotherdynamic}\\
		\epsilon\frac{\partial x_2}{\partial \theta}&=\zeta_1x_1-x_2, 
		\label{x22anotherdynamic}		
	\end{align}
\end{subequations}	
where $\epsilon=\frac{k_1k_2}{(k_{-1}+k_2)^2}, K_3=\frac{k_{-3}}{k_2}, K_1=\frac{k_{-1}}{k_2}$. By considering $\epsilon$ is very small quantity, we can substitute $x_1, x_2, e$ in eq. \eqref{zeta1anotherdynamic} and \eqref{zeta2anotherdynamic} by their steady state values. Thus, eq. \eqref{zeta1anotherdynamic} and \eqref{zeta2anotherdynamic} become 
\begin{subequations}
	\begin{align}
		\frac{\partial \zeta_1}{\partial \theta}&=z-\frac{\zeta_1\zeta_2^2}{1+\zeta_2^2(1+\zeta_1)}\label{zeta1dynamic}\\
		\frac{\partial \zeta_2}{\partial \theta}&=\alpha_2[\frac{\zeta_1\zeta_2^2}{1+\zeta_2^2(1+\zeta_1)}-X_2\zeta_2], 
		\label{zeta2dynamic}		
	\end{align}
\end{subequations}

For further simplification we introduce following rescaled quantities, $\tau=X_2^2z^{-2}\theta; x=X_2^{-1}z\alpha_2\zeta_1; y=X_2^{-1}z\zeta_2; \nu=X_2^{-3}z^4\alpha_2; \omega=X_2^{-1}z^2\alpha_2; \kappa=z{X_2}^{-1}  \alpha_2 $ in eq. \eqref{zeta1dynamic} and  \eqref{zeta2dynamic}, and obtain following system,
\begin{subequations}
	\begin{align}
		\frac{\partial x}{\partial \tau}&=\nu-\frac{xy^2}{1+\frac{z}{\nu} y^2(\kappa+x)}\label{xdynamic}\\
		\frac{\partial y}{\partial \tau}&=\frac{x y^2}{1+\frac{z}{\nu} y^2(\kappa+x)}-\omega y.
		\label{ydynamic}		
	\end{align}
\end{subequations}
Due to slow glycolytic flux during self-oscillation i.e. for $z<<1$, we can write ${1+\frac{z}{\nu} y^2(\kappa+x)}=1$. Finally, the simplified form of the Selkov model system can be written as, 
\begin{equation}
\begin{aligned}
\frac{\partial x}{\partial \tau}&=\nu-xy^2 \\
\frac{\partial y}{\partial \tau}&=xy^2-\omega y. 
\label{reselkov}
\end{aligned}
\end{equation}

\subsection{Chemostated Selkov Model}
For the nonequilibrium thermodynamic representation of the chemical reaction network, all the elementary chemical reactions must be reversible. However, as mentioned earlier, the elementary reaction of $\rho=2$ in eq. \eqref{crn} in Selkov scheme is  irreversible. An equivalent completely reversible description based on the Selkov model can be written as \citep{reversibleselkoov}

\begin{equation}
\begin{aligned}
\rho&=1:&\ch{A&<=>[\text{k_1}][\text{k_{-1}}]S} \\
\rho&=2:&\ch{S + 2 P &<=>[\text{k_2}][\text{k_{-2}}] 3 P}\\
\rho&=3:& \ch{P&<=>[\text{k_{3}}][\text{k_{-3}}]
	B}\\ 
\end{aligned}
\label{revcrn}
\end{equation}
with S and P are the ATP and ADP concentrations, respectively. This reversible reaction network would be convenient in connecting kinetic and thermodynamic description. Here $\{S, P\}\in I$ are two intermediate species having dynamic concentration and  $\{A , B\}\in C$ are externally controllable chemostatted species. Considering that reverse reaction rates are very small relative to the forward reaction rates, we can write the dynamical equation of eq. \eqref{revcrn} as 
\begin{equation}
\begin{aligned}
\frac{\partial s}{\partial t}&=k_1a-k_2 s p^2\\
\frac{\partial p}{\partial t}&=k_2s p^2-k_3p
\label{revdynamic}
\end{aligned}
\end{equation}
By introducing scaled variables, $\tau=\frac{k_2t}{c_1^2}, \nu=\frac{c_1^3k_1a}{k_2},  x=c_1s, y=c_1p, \omega=\frac{c_1^2k_3}{k_2}$ in eq. \eqref{revdynamic}, we would have the same set of equations as the eq. \eqref{reselkov}. Here $c_1$ is an arbitrary constant. We would use the parameter, $\nu$ as the control parameter of the system keeping another parameter, $\omega$ at a fixed value. 

\section{\label{LSA}Dynamical Stability of the System with Diffusion}

When the spatial aspect of the system is not considered, we can only have Hopf instability with uniform oscillation with wavenumber $q=0$ in the model. In a more general case with a nonzero finite wavenumber, the reaction-diffusion model of glycolysis also admits traveling waves.

A unique steady-state value of the eq. \eqref{reselkov} that satisfies $\dot{x}=\dot{y}=0$ is,
$x_0=\frac{\omega^2}{\nu}, y_0=\frac{\nu}{\omega}$. For linear stability analysis at the steady-state value $(x_0,y_0)$, one needs to consider the Jacobian matrix of the model,
\begin{gather}
	\mathcal{J}=
	\begin{pmatrix}   
		-{y_0}^2 & -2{x_0y_0} \cr
		{y_0}^2& 2{x_0y_0}-\omega\cr
	\end{pmatrix}.\label{jaco}                 
\end{gather}
Elements of the Jacobian matrix, $\mathcal{J}$ are the following,

$J_{11}=-{y_0}^2$, $J_{12}=-2{x_0y_0}, J_{21}={y_0}^2$, $J_{22}=2{x_0y_0}-\omega.$

Determinant and  trace  of the Jacobian, $\mathcal{J}$ are $det(\mathcal{J})=\frac{\nu^2}{\omega}$ and
$Tr(\mathcal{J})=\omega-(\frac{\nu}{\omega})^2$,  respectively.  Eigenvalues, $\lambda$ of $\mathcal{J}$ are given by the characteristic equation, 
$\lambda^2-Tr(\mathcal{J})\lambda+det(\mathcal{J})=0.$
Hence eigenvalues in terms of determinant and trace are,
\begin{equation}
\lambda_{\pm}=\frac{Tr(\mathcal{J})\pm\sqrt{Tr(\mathcal{J})^2-4det(\mathcal{J})}}{2}.
\label{eigenvalue}
\end{equation}
As chemical parameters are real quantities,  eigenvalues of the system at stable steady-state are complex conjugate pair $\lambda_{\pm}=\lambda_{r}\pm i\lambda_{i}$.
At the onset of Hopf instability, $Tr(\mathcal{J})=0$, i.e.,  $J_{11}+J_{22}=0$ and it leads to  a critical value of the control parameter as,  $\nu_{cH}=\omega\sqrt{\omega}.$ Therefore, at the onset of the Hopf instability, the determinant is $det(\mathcal{J})=\omega^2$ and eigenvalues are $\lambda_{\pm}=\pm i \omega$. The critical frequency of the Hopf instability, $f_{cH}$ will be the imaginary part of the eigenvalue, $\lambda$  at the onset of instability and hence the period of the limit cycle near the onset of instability, $\nu_{cH}$ is approximately, $T=\frac{2\pi}{f_{cH}}$, where $f_{cH}=\omega$. Now the critical eigenvector, $U_{cH}$, corresponding to the eigenvalue, $\lambda=i \omega$ at the onset of Hopf instability is,
\begin{equation}
U_{cH}=\begin{pmatrix}
1-i\cr
-1
\end{pmatrix}.
\end{equation}

Now in the presence of diffusion, the reaction diffusion equation of the Selkov model in one spatial dimension $r\in [0,l]$ can be expressed from eq. \eqref{reselkov} as
\begin{equation} 
\begin{aligned}
\frac{\partial x}{\partial \tau}&=\nu-xy^2+D_{11}x_{rr}+D_{12}y_{rr} \\
\frac{\partial y}{\partial \tau}&=xy^2-\omega y+D_{21}x_{rr}+D_{22}y_{rr} 
\label{ddynamic}
\end{aligned}
\end{equation}
in which $D_{11}$ , $D_{22}$ are self-diffusion coefficients corresponding to intermediate species, $x$ and $y$ respectively and $D_{12}$ , $D_{21}$ are cross-diffusion coefficients of $x$  and $y$. Diffusion coefficients can have concentration dependence. However, we have considered here constant self- and cross-diffusion coefficients for simplicity. 

In the presence of diffusion, Jacobian $\mathcal{J}$ becomes
\begin{gather}
	\mathcal{J_D}=\mathcal{J}-q^2\mathcal{D}\nonumber\\
	=\begin{pmatrix}   
		-{y_0}^2 & -2{x_0y_0} \cr
		{y_0}^2& 2{x_0y_0}-\omega\cr
	\end{pmatrix}
	- q^2
	\begin{pmatrix}
		D_{11}&D_{12}\cr
		D_{21}&D_{22}\cr
	\end{pmatrix}
	\label{jacod}                 
\end{gather}
where we have applied a Fourier transform $g(r,t)\rightarrow g(q,t)$, with $q$ being the wavenumber. Now the trace of the $\mathcal{J_{D}}$ is 
$Tr(\mathcal{J_{D}})=Tr(\mathcal{J})-q^2Tr(\mathcal{D})=\omega-(\frac{\nu}{\omega})^2-(D_{11}+D_{22})q^2$ and the determinant of $\mathcal{J_{D}}$ is a quadratic equation of $q^2$
\begin{eqnarray}
det(\mathcal{J_D})=det(\mathcal{D})q^4-[D_{11}J_{22}+D_{22}J_{11}\nonumber \\
-D_{12}J_{21} 
-D_{21}J_{12}]q^2+det(\mathcal{J}),
\label{deter}
\end{eqnarray}
in which $det(\mathcal{J})=\frac{\nu^2}{\omega}$ is the determinant of  $\mathcal{J}$ and $det(\mathcal{D})$ is the determinant of the matrix containing diffusion coefficients. Eigenvalues, $\lambda$ of $\mathcal{J_{D}}$ are obtained by the characteristic equation,  
$\lambda^2-Tr(\mathcal{J_{D}})\lambda+det(\mathcal{J_{D}})=0.$
Hence eigenvalues can be expressed only in terms of determinant and trace as
\begin{equation}
\lambda_{\pm}=\frac{Tr(\mathcal{J_{D}})\pm\sqrt{Tr(\mathcal{J_{D}})^2-4det(\mathcal{J_D})}}{2}.
\label{eigenvaluealt}
\end{equation}
The stability criterion would demand both of these eigenvalues have to be negative, and thus in terms of trace and determinant, this implies $Tr(\mathcal{J_{D}}) <0$ and $det(\mathcal{J_D}) >0$.
The existence of the traveling wave in the presence of diffusion demands that $Tr(\mathcal{J_{D}})=0$ and $det(\mathcal{J_D})>0$. 

Now exploiting the $Tr(\mathcal{J_{D}})=0$ condition, the critical value of the control parameter, $\nu$ can be specified as,
\begin{equation}
\nu_{ctw}=w\sqrt{w-(D_{11}+D_{22})q^2}. 
\end{equation} 
The wavenumber $q$ has to follow $q=\frac{2n\pi}{l}$ according to periodic boundary conditions in the finite domain of size $l$. Here $n$ is an integer and  specifies the number of oscillations within the region of interest. The condition $det(\mathcal{J_D})>0$ for generating traveling wave imposes a restriction on the  wavenumber selection. Hence the wavenumber $q$ in this model needs to satisfy the condition obtained from \eqref{deter},
\begin{equation}
det(\mathcal{D})q^4-[D_{11}-D_{22}-D_{12}+2D_{21}]wq^2+w^2>0.
\label{wncon}
\end{equation}  
      
\section{\label{selampli}Amplitude Equation in the Presence of Cross Diffusion}

The amplitude of a dynamical system is generally a complex quantity that demonstrates features akin to the order parameter in a phase transition. We have used the KB averaging method\citep{krylov1949introduction} to find out the magnitude and phase dynamical equation of the extended Selkov model in the presence of both self and cross-diffusion. In this KB method, the slowly varying magnitude and phase allow us to treat the instantaneous amplitude equation and the averaged amplitude equation on an equal footing. 

Initially, two new variables, i. e., the total concentration of intermediate species $z=x+y$ and total flux of $u=\nu-\omega y$, have been introduced to rewrite eq. \eqref{reselkov} of the Selkov model in the following form,
\begin{equation}\begin{aligned}  \dot{z}&=u \\  \dot{u}&=-\omega(u-\nu)-\omega^{-2}(\omega z+u-\nu)(u-\nu)^2  \label{eq:2}
\end{aligned}
\end{equation} 
The steady state solution of the new set of differential equations is  $u_s=0$ and $z_s=\frac{\omega^2}{\nu}+\frac{\nu}{\omega}$. From now on, we would use the notation, $t$ in the place of $\tau$ to denote time. Now setting up another new variable as $\zeta=z-z_s$ will shift the fixed point of the system to the origin. With the aid of $u$ and $\zeta$ it is possible to represent  \eqref{eq:2} as a single second-order equation akin to the generalized Rayleigh equation\citep{LAVROVA2009127, SGDSR, Rayleigh1945TheSound},
\begin{equation}
\ddot{\zeta}+\Omega^2\zeta=\lambda[2(1+c_1u-c_2u^2)u-\frac{\Omega^2}{\lambda}(\nu^{-2}u^2-2\nu^{-1} u)\zeta]
\label{eq:3}
\end{equation}
where  $\Omega=\frac{\nu}{\sqrt{\omega}}, \lambda=\frac{\omega-\omega^{-2}\nu^2}{2}, c_1=\frac{(2\omega^{-2}\nu-\frac{\omega}{\nu})}{2\lambda}, c_2=\frac{\omega^{-2}}{2\lambda}$.
By inserting $2(1+c_1u-c_2u^2)u-\frac{\Omega^2}{\lambda}(\nu^{-2}u^2-2\nu^{-1} u)\zeta=h$ in the eq. \eqref{eq:3}, we obtain
\begin{equation}
\ddot{\zeta}+\Omega^2\zeta=\lambda h. 
\label{zeta}
\end{equation}
Now for the reaction-diffusion representation of the Selkev model, eq. \eqref{ddynamic} comprising of both self- and cross-diffusion coefficients, we can extend  eq. \eqref{zeta} in the following way
\begin{eqnarray}
\ddot{\zeta}+\Omega^2\zeta=\lambda h+(D_{22}+D_{12}-D_{11}-D_{21})\dot{u}_{rr}\nonumber\\
+(D_{22}+D_{12})\dot{\zeta}_{rr}
+(D_{11}-D_{12})u_{rr}-D_{12}\zeta_{rr}. \label{dzeta}
\end{eqnarray} 
For $\lambda$ being infinitesimal, the eq. \eqref{dzeta} would accept simple harmonic function like solutions,
\begin{subequations}
	\begin{align}
		\zeta(r,t)&=\mathcal{A}(r,t)\cos(\Omega t+\phi(r,t))\label{sh1}\\
		u(r,t)&=\dot{\zeta}(r,t)=-\Omega \mathcal{A}(r,t)\sin(\Omega t+\phi(r,t))\label{sh2}  
	\end{align}
\end{subequations}
with slowly varying  amplitude, $\mathcal{A}$ and phase, $\phi$ during fast oscillations. Finally, with the aid of eq. \eqref{sh1} and \eqref{sh2}, we acquire the dynamical equations of amplitude and phase, respectively as,
\begin{eqnarray}
\dot{\mathcal{A}}=-\frac{1}{\Omega}[\lambda h-\Omega^2(D_{22}+D_{12}+\frac{D_{12}}{\Omega^2}-D_{11}\nonumber \\
-D_{21})\zeta_{rr}+(D_{22}+D_{11})u_{rr}] \sin(\Omega t+\phi),
\label{amp}\\
\dot{\Phi}=\frac{1}{\Omega \mathcal{A}}[\lambda h-\Omega^2(D_{22}+D_{12}+\frac{D_{12}}{\Omega^2}-D_{11}\nonumber \\
-D_{21})\zeta_{rr}+(D_{22}+D_{11})u_{rr}]\cos(\Omega t+\phi).
\label{phs} 
\end{eqnarray} 
Now by taking the average, amplitude and phase equations of the Selkov reaction-diffusion model in the presence of cross diffusion are obtained as,
\begin{subequations}
	\begin{align}
		\dot{\mathcal{A}}=\mathcal{A}\lambda-p_{1}\frac{3\lambda c_{2}\Omega^2}{4}A^3-\frac{\Omega}{2}(D_{22}+D_{12}+\frac{D_{12}}{\Omega^2}-D_{11}\nonumber \\ -D_{21})(2\mathcal{A}_{r}\phi_{r}+\phi_{rr}\mathcal{A})+\frac{(D_{11}+D_{22})}{2}(\mathcal{A}_{rr}-\mathcal{A}\phi_{r}^2), \label{kbamp}\\
		\dot{\Phi}=-p_{2}\frac{\Omega^3}{8\nu^2}\mathcal{A}^2-\frac{(D_{11}+D_{22})}{2} (\frac{2\mathcal{A}_{r}\phi_{r}}{\mathcal{A}}+\phi_{rr})-\frac{\Omega}{2}(D_{22}\nonumber \\
		+D_{12}+\frac{D_{12}}{\Omega^2}-D_{11}-D_{21})(\frac{\mathcal{A}_{rr}}{\mathcal{A}}-\phi_{r}^2).
		\label{kbphase}
	\end{align}
\end{subequations}
Here, correction factors,  $p_1=\frac{c_2}{c_1}$ and $p_2=\frac{2c_1}{\nu c_2}$ are introduced in eq. \eqref{kbamp} and \eqref{kbphase} to include the modification in radius and phase of cycle owing to unidirectional acceleration from unstable steady state\citep{Lavrova2009PhaseInflux}. 

\subsection{\label{sec:hae}Complex Ginzberg-Landau Equation Near Hopf Onset: Amplitude and Phase Equations}   

Near the onset of Hopf instability, the lowest-order amplitude equation, Complex Ginzburg-Landau equation(CGLE)\citep{aranson2002world, Cross2009PatternSystems} properly reflects the dynamics of the patially extended nonlinear oscillatory models. The unscaled form of CGLE can be represented as,
\begin{equation}
\frac{\partial Z}{\partial t}=\lambda Z -(\beta_{r}-i\beta_{i})\mid Z \mid ^2 Z+(\alpha_{r}+ i \alpha_{i})\partial_{r}^2 Z.
\label{cgl}
\end{equation}
Assuming the same velocity for all the traveling waves and introducing a comoving coordinate as $r=r-velocity\times t $, we can utilize the same amplitude equation of the form of eq.  \eqref{cgl} for the traveling waves.
  
By inserting  $Z=\mathcal{A}\exp({i \phi})$ in eq. \eqref{cgl} and separating real and imaginary parts we obtain,
\begin{subequations}
	\begin{align}
		\frac{\partial \mathcal{A}}{\partial t}&=\lambda \mathcal{A} -\beta_{r}\mathcal{A}^3-\alpha_{i}(2\mathcal{A}_{r}\phi_{r}+\phi_{rr}\mathcal{A})+\alpha_{r}(\mathcal{A}_{rr}-\mathcal{A}\phi_{r}^2), \label{ampd}\\
		\frac{\partial \phi}{\partial t}&=\beta_{i}\mathcal{A}^2+\alpha_{r}(\frac{2\mathcal{A}_{r}\phi_{r}}{\mathcal{A}}+\phi_{rr})+\alpha_{i}(\frac{\mathcal{A}_{rr}}{\mathcal{A}}-\phi_{r}^2). \label{phased}
	\end{align}
\end{subequations}        
Comparing amplitude and phase equations  \eqref{ampd} and  \eqref{phased} deduced from CGLE with eq. \eqref{kbamp} and \eqref{kbphase} derived by the KB method, we get the following coefficients: 
$\beta_{r} =p_{1}\frac{3\lambda c_{2}\Omega^2}{4}$,  $\beta_{i} =-p_{2}\frac{ \Omega^3}{8\nu^2}$, $\alpha_{r} =-\frac{(D_{11}+D_{22})}{2}$, $\alpha_{i} =-\frac{\Omega}{2}(D_{22}+D_{12}+\frac{D_{12}}{\Omega^2}-D_{11}-D_{21})$. With the help of  scaled variables $\mathcal{A}=\frac{\mathcal{A}}{\sqrt{\beta_{r}}}$ and $r=\frac{r}{\sqrt{\alpha_{r}}}$, we can represent eq. \eqref{ampd} and \eqref{phased} as 
\begin{subequations}
	\begin{align}
		\frac{\partial \mathcal{A}}{\partial t}&=\lambda \mathcal{A} -\mathcal{A}^3-\alpha(2\mathcal{A}_{r}\phi_{r}+\phi_{rr}\mathcal{A}) +(\mathcal{A}_{rr}-\mathcal{A}\phi_{r}^2), \label{samp}\\
		\frac{\partial \phi}{\partial t}&=\beta \mathcal{A}^2+(\frac{2\mathcal{A}_{r}\phi_{r}}{\mathcal{A}}+\phi_{rr})+\alpha(\frac{\mathcal{A}_{rr}}{\mathcal{A}}-\phi_{r}^2), \label{sphase}
	\end{align}
\end{subequations}
and the corresponding normal form of CGLE\citep{Nicolis1995IntroductionScience, Walgraef1997ThePatterns, Cross2009PatternSystems} in spatially extended system as 
\begin{equation}
\frac{\partial Z}{\partial t}=\lambda Z -(1-i\beta)\mid Z \mid ^2Z+(1+i \alpha)\partial_{r}^2 Z.
\label{ncgle}
\end{equation}
Coefficients in normal form of CGLE are given by $\alpha=\frac{\alpha_{i}}{\alpha_{r}}=\frac{\Omega(D_{22}+D_{12}+\frac{D_{12}}{\Omega^2}-D_{11}-D_{21})}{(D_{11}+D_{22})}$ and 
$\beta=\frac{\beta_{i}}{\beta_{r}}=-\frac{p_2}{p_1}\frac{\sqrt{\omega}\omega}{3\nu }.$
In the case of Hopf instability, it is apparent that only the coefficient $\alpha$ explicitly depends on both the self- and cross-diffusion terms. 

For large $r$, the normal form of the CGLE eq. \eqref{ncgle} has an asymptotic solution of simple plane wave for nonlinear oscillations,
\begin{equation}
Z=\mathcal{A}\exp{i(\omega_{0}t+qr)}
\label{trialeq}
\end{equation}
with $\omega_{0}$ is the shift in frequency from the critical frequency $\omega_{cH}$. Here, $q$ is a unique wave number selected by the unique spiral frequency. With the help of eq. \eqref{trialeq}, we obtain  from eq. \eqref{ncgle}
\begin{subequations}
	\begin{align}
		\mathcal{A}^2&=\lambda-q^2\label{steadyA}\\ \omega_{0}&=\omega_{q}-\omega_{cH}=\beta \mathcal{A}^2-\alpha q^2=\beta \lambda-(\beta+\alpha) q^2\label{disperionnonlinear}.
	\end{align}
\end{subequations}
\begin{figure*}
	\centering 
	\subfigure[\label{fig.wavenumbervsalpha}]{\includegraphics[width=0.455\linewidth]{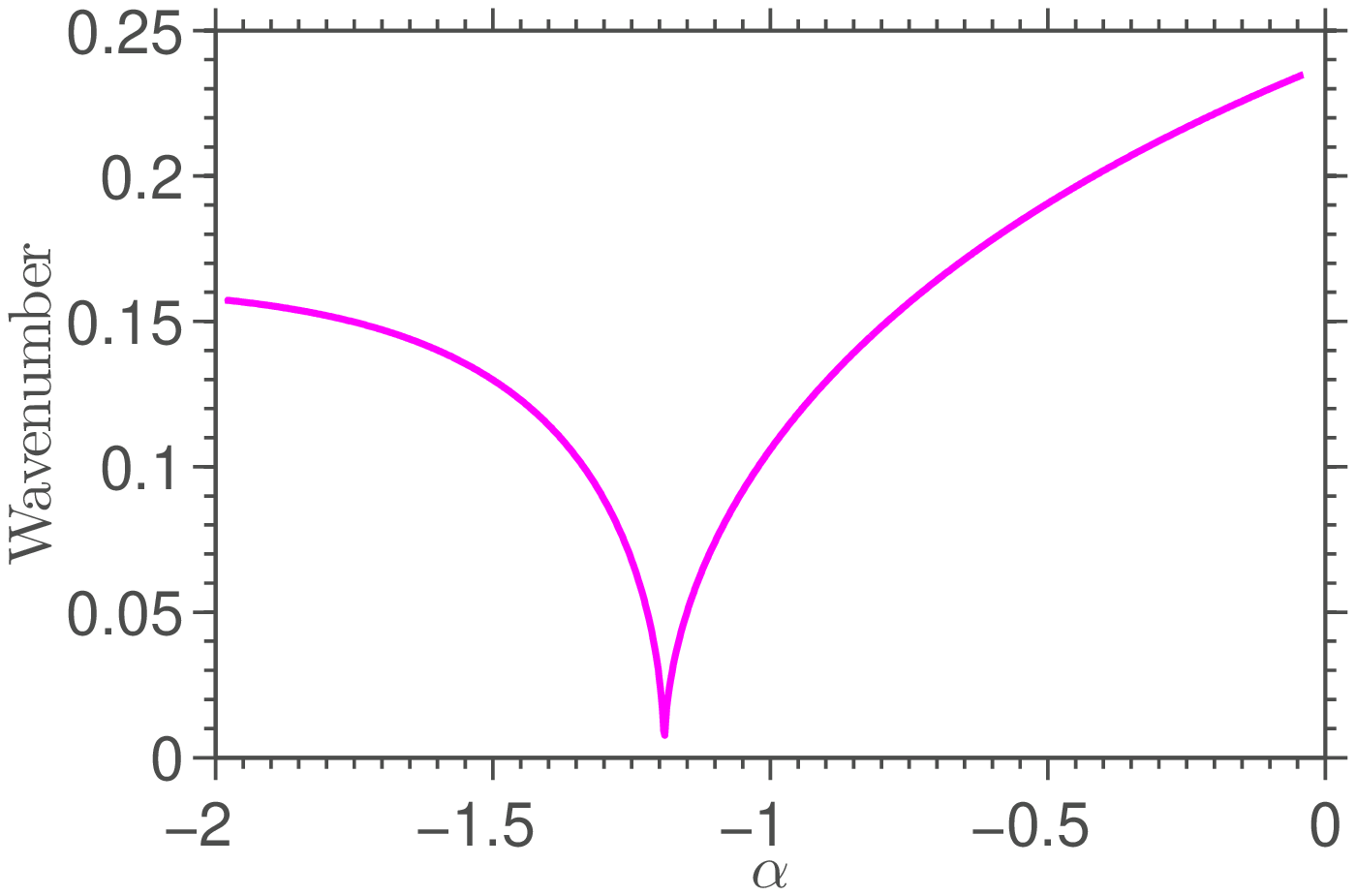}}\hfill
	\subfigure[\label{fig.magnitudevsalpha}]{\includegraphics[width=0.525\linewidth]{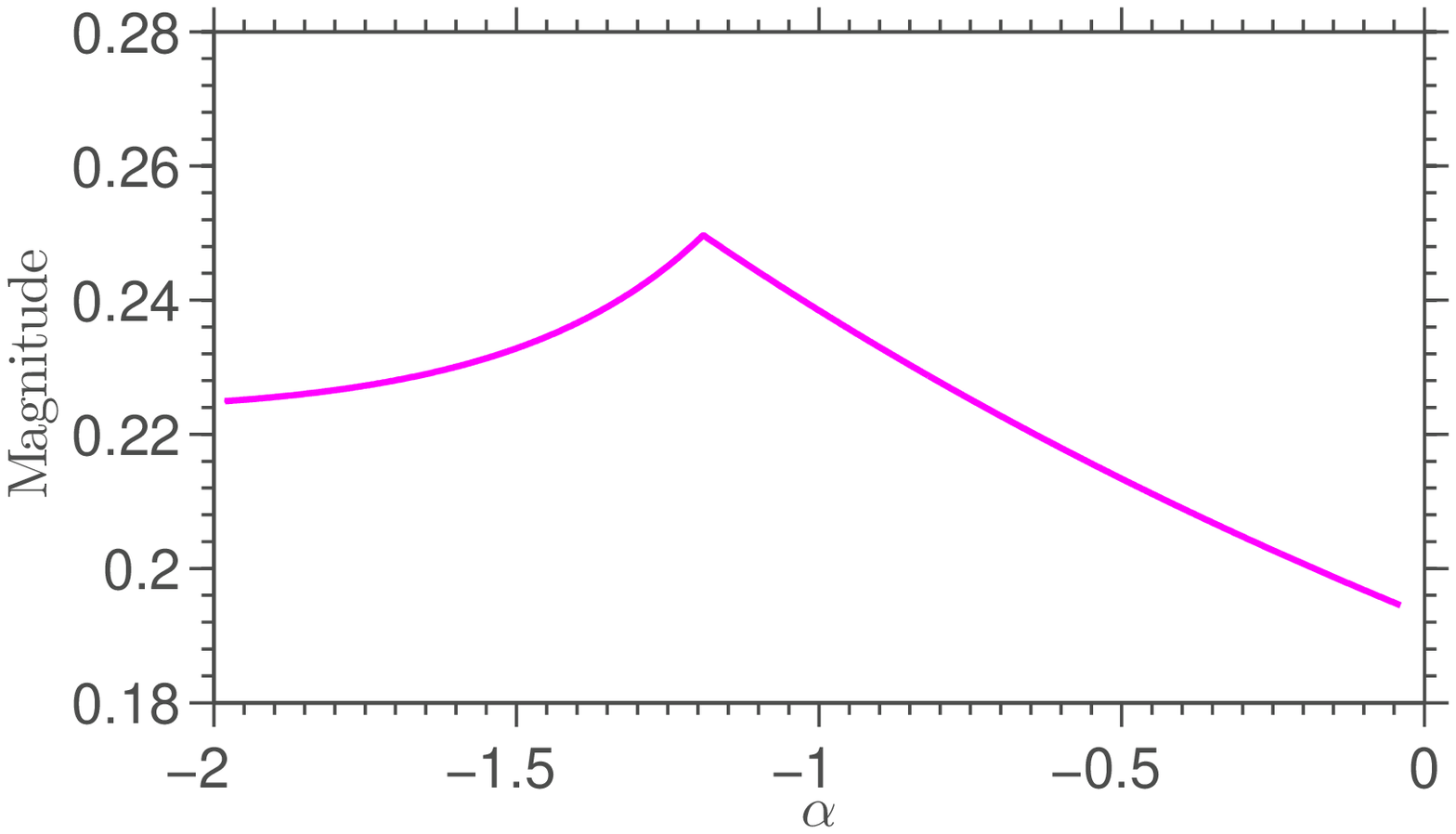}}
	\label{dynamicdiffusion}
	\caption{ Wavenumber against the amplitude equation coefficient $\alpha$ is shown in FIG. \ref{fig.wavenumbervsalpha}. FIG. \ref{fig.magnitudevsalpha} illustrates variation of the magnitude of the complex amplitude with respect to same coefficient, $\alpha$. Both the figures are obtained by varying cross-diffusion coefficient, $D_{12}$ from $-0.00001$ to $-0.0005$ and $D_{21}$ from $0.00001$ to $0.0005$ while other parameters are fixed i.e. $D_{11}=D_{22}=0.00051$, $\nu=2.45$ and $w=2$.}
\end{figure*}
Here the bulk frequency for the system or the frequency of uniform oscillation is obtained by inserting $q=q_{cH}=0$ in \eqref{disperionnonlinear}. In general, the phase and  group velocities can be expressed by  $v_p=\frac{\omega_{q}}{q}$  and  $v_g=\frac{\partial \omega_{q}}{\partial q}=-2(\beta+\alpha)q$, respectively and they can have different sign.

For slow time variation of amplitude, we can set $\mathcal{A}$ to its steady state variation. Therefore we can write from eq. \eqref{samp}, 
\begin{equation}
\mathcal{A}^2=\lambda -\alpha(\frac{2\mathcal{A}_{r}\phi_{r}}{\mathcal{A}}+\phi_{rr}) +(\frac{\mathcal{A}_{rr}}{\mathcal{A}}-\phi_{r}^2). \label{ssamp}
\end{equation}
The phase dynamical equation, eq. \eqref{sphase} contains the space derivatives of $\mathcal{A}$ and due to long range phase variation we would remove the higher space derivatives of $\mathcal{A}$ in final equation. Thus the steady state amplitude, eq. \eqref{ssamp} can be simplified to  
\begin{equation}
\mathcal{A}^2=\lambda -\alpha \phi_{rr} -\phi_{r}^2. \label{overssamp}
\end{equation}
By inserting eq. \eqref{overssamp} into eq. \eqref{sphase}, we obtain the nonlinear phase dynamical equation as 
\begin{equation}
\frac{\partial \phi}{\partial t}=\beta\lambda +(1-\alpha\beta)\phi_{rr}-(\alpha+\beta)\phi_{r}^2. \label{nonlinearphase}	
\end{equation}	
Now introducing new phase variable, $\psi=\phi-\beta\lambda t$, we acquire from \eqref{nonlinearphase}	
\begin{equation}
\frac{\partial \psi}{\partial t}=(1-\alpha\beta)\psi_{rr}-(\alpha+\beta)\psi_{r}^2. \label{reducednonlinearphase}
\end{equation}

The exchange between inward and outward rotating spiral is associated with the criterion, $(\alpha+\beta)<0$, whereas the  Newell criterion, $(1-\alpha\beta)<0$ gives rise to the Benjamin-Feir(BF) instability\citep{benjamin1967instability, Cross2009PatternSystems} from the uniform oscillation. The BF instability, a long-wave side-band instability, was first identified in deep-water waves  \citep{benjamin1967instability}. Due to the onset of BF instability, the wavenumber, and frequency of a previously uniform traveling wave become irregular.

 Now with the aid of derivation with respect to space and setting $\psi_r=u$, we obtain the following equation similar to the Burger's equation form, \eqref{reducednonlinearphase} 
\begin{equation}
\frac{\partial u}{\partial t}=(1-\alpha\beta)u_{rr}-(\alpha+\beta)2u u_{r}. \label{burger}
\end{equation}
 Application of Cole-Hopf transformation, $\psi=-[\frac{1-\alpha \beta}{\alpha+\beta}]\ln{\chi}$ to \eqref{reducednonlinearphase} will transform it into a linear equation,
\begin{equation}
\frac{\partial \chi}{\partial t}=(1-\alpha\beta)\chi_{rr}. \label{linearburger}
\end{equation}
As phase is a real variable, the trial solution of linear dynamical equation, eq. \eqref{linearburger} can be considered as $\chi= G(t)\exp{(\alpha+\beta)qr}$ and inserting the trial solution into eq. \eqref{linearburger}, we would have $G(t)=G_0\exp{(1-\alpha \beta)(\alpha+\beta)^2q^2t}$. Therefore a simple solution to eq. \eqref{linearburger} is $\chi= G_0\exp{((1-\alpha \beta)(\alpha+\beta)^2q^2t+(\alpha+\beta)qr)}$.
Accordingly, the expression of the phase of the system  is
 \begin{equation}
 \phi=\beta \lambda t-\frac{1-\alpha \beta}{\alpha+\beta}[\ln{G_0}+(1-\alpha \beta)(\alpha+\beta)^2q^2t+(\alpha+\beta)qr].\label{mainphase}
 \end{equation}
It should be noted that modification of the system's frequency will lead to the change of wavenumber, $q$ according to the nonlinear dispersion relation in eq. \eqref{disperionnonlinear}.

\subsection{\label{pws} Stability of Plane Wave}  

To test the stability of the asymptotic plane wave solution of the CGLE, we can afford a small perturbation about the nonlinear wave state. Thus the perturbed plane wave would have the following form 
\begin{equation}
Z=(\mathcal{A}+A_{per})\exp{i(qr+\omega_0 t)}.
\label{perwave}
\end{equation} 
Here $A_{per}$ can be expressed in terms of complex growth rate, $\sigma$ in the following way
\begin{equation}
A_{per}=A_{+}\exp{(iKr+\sigma t)}+\tilde{A_{-}}\exp{(-iKr+\tilde{\sigma} t)}
\label{perterm}
\end{equation} 
with $\tilde{A}$ and $\tilde{\sigma}$ are complex conjugation of $A$ and $\sigma$ and $K$ corresponds to different perturbation modes. Now inserting eq. \eqref{perwave} into  eq. \eqref{ncgle} and neglecting higher order of the perturbation, we arrive at following equation
\begin{eqnarray}
\frac{\partial A_{per}}{\partial t}+i\omega_0(\mathcal{A}+A_{per})=\lambda(\mathcal{A}+A_{per})
-(1-i\beta)(\mathcal{A}^3\nonumber\\
+2\mathcal{A}^2A_{per}+\mathcal{A}^2\tilde{A}_{per})+
(1+i\alpha)({\partial_r}^2 A_{per}+2iq \partial_r A_{per}\nonumber\\
-q^2(\mathcal{A}+A_{per})).\nonumber\\
\label{permiddle}
\end{eqnarray}
Now substituting eq. \eqref{perterm} into eq. \eqref{permiddle} and rearranging in terms of functions $\exp{(iKr+\sigma t)}$ and  $\exp{(-iKr+\tilde{\sigma t})}$, we have 
\begin{eqnarray}
[A_{+}\sigma +i\omega_0 A_{+}-\lambda A_{+}+(1-i\beta)(2\mathcal{A}^2A_{+}+
\mathcal{A}^2A_{-})\nonumber\\
+(1+i\alpha)(K
+q)^2A_+]\exp{(iKr+\sigma t)}\nonumber\\
+[\tilde{A}_{-}\tilde{\sigma} +i\omega_0 \tilde{A}_{-}-\lambda \tilde{A}_{-}+(1-i\beta)(2\mathcal{A}^2\tilde{A}_{-}+
\mathcal{A}^2\tilde{A}_{+})\nonumber\\
+(1+i\alpha)(K-q)^2\tilde{A}_-]\exp{(-iKr+\tilde{\sigma t})}\nonumber\\
+i\omega_0\mathcal{A}-\lambda\mathcal{A}+(1-i\beta)\mathcal{A}^3+(1+i\alpha)q^2\mathcal{A}=0.\nonumber\\
\label{perarrange}
\end{eqnarray}
Equating the coefficients of functions $\exp{(iKr+\sigma t)}$ and  $\exp{(-iKr+\tilde{\sigma t})}$ to $0$, we obtain a homogeneous system comprising of two linear equations 
\begin{subequations}
\begin{align}
(\sigma+(1-i\beta)\mathcal{A}^2+(1+i\alpha)(K^2+2qK))A_+ \nonumber\\
+(1-i\beta)\mathcal{A}^2 A_-
=0	 \label{persub1} \\
(\sigma+(1+i\beta)\mathcal{A}^2+(1-i\alpha)(K^2-2qK))A_-  \nonumber\\ +(1+i\beta)\mathcal{A}^2 A_+=0.	\label{persub2} 
\end{align}
\end{subequations}	
We can represent eq. \eqref{persub1} and eq. \eqref{persub2} as
	\begin{gather}H
		\begin{pmatrix}
			A_+\cr
			A_-
		\end{pmatrix}=
		0
		\end{gather}
		where 
\[H=
\begin{bmatrix}
		\sigma+(1-i\beta)\mathcal{A}^2\\+(1+i\alpha)(K^2+2qK)	 & (1-i\beta)\mathcal{A}^2 \cr
		& \cr
(1+i\beta)\mathcal{A}^2	 & \sigma+(1+i\beta)\mathcal{A}^2\cr
&+(1-i\alpha)(K^2-2qK)	
\end{bmatrix}
.\]		
By setting $det(H)=0$, we can find the characteristic equation for $\sigma$ and solving the characteristic equation, we get the most positive growth rate $\sigma$ as, 
\begin{equation}
\begin{split}
\sigma=-\mathcal{A}^2-K^2-2i\alpha q K 
\\+\sqrt{(1+\beta^2)\mathcal{A}^4-(\alpha K^2-\beta \mathcal{A}^2-2i q K )^2}.
\end{split}
\label{characsigma}
\end{equation} 
Now to explore the long wavelength behavior, we have expanded eq. \eqref{characsigma} around $K=0$,
\begin{equation}
\sigma=-2iq(\alpha+\beta)K+[\frac{(2q^2(1+\beta^2)+\beta\alpha \mathcal{A}^2)}{\mathcal{A}^2}-1]K^2+O(K^3).
\label{sigmataylor}
\end{equation}
We seek the threshold of stable  wavenumber above which traveling wave shows instability. The condition $\partial_{KK}\sigma=0$ sets a boundary between stable and unstable wavenumber. Therefore, at the onset of instability one can find
\begin{equation}
q^2= \frac{\lambda(1-\alpha \beta)}{2\beta^2-\alpha \beta+3}.
\label{selected_wavenumber}
\end{equation} 

Thus wavenumber lies within the band set by eq. \eqref{selected_wavenumber} near the critical wavenumber of the Hopf instability would be the allowed wavenumber of the traveling wave. Unlike the equilibrium system, the wavenumber selection is a significant problem to be addressed in a finite system far from equilibrium \citep{Cross2009PatternSystems}. We can have different wavenumbers corresponding to control parameter values based on different boundary conditions, dynamical processes, perturbations, and methodologies. Here we have considered a finite system with periodic boundary conditions.  The wavenumber $q$ has to satisfy
$q=\frac{2n\pi}{l}$ where $n$ is an integer to fit the domain length, $l$, and periodic boundary conditions. Therefore wavenumber is quantized, and we have a discrete set of possible wavenumbers of the plane waves as the control parameter of the system is tuned. Now the perturbation wavenumber, $K$, also needs to fit in the finite domain with periodic boundary conditions, and thus we can write, allowed wavenumber, $q\pm K=\frac{2m\pi}{l}$. Here $m$ is also an integer and generally $m\neq n$. The perturbation wavenumber can be kept at a minimum finite value by maintaining $|m-n|=1$ near the critical wavenumber of the Hopf instability. Thus we obtain a discontinuous change in allowed wavenumbers for the control parameter. Now within the BF instability regime, plane waves are linearly unstable. However, due to the convective nature of the instability, linearly unstable waves can hold some physical relevance. Near the onset of BF instability, $q$ approaches zero. For $q=0$, eq. \
 eqref{characsigma} results in following equation,

\begin{equation}
\sigma_{q=0}=-\lambda-K^2+\sqrt{\lambda^2-\alpha^2K^4+2\alpha \beta K^2\lambda}.
\label{sigmaqzero}
\end{equation}
Now expanding eq. \eqref{sigmaqzero} for $K\rightarrow0$ and then setting the Taylor expansion of  $\sigma_{q=0}$ to zero, we arrive at following expression 
\begin{equation}
K_c^2=\frac{2\lambda(\alpha\beta-1)}{\alpha^2(1+\beta^2)}.
\label{crtick}
\end{equation}
For the control parameter value extremely near the BF instability onset and within the BF instability regime, linearly unstable modulated waves with the discrete allowed wavenumbers obtained from the band $|K|<K_c$ persist in the system. The effect of the finite domain size on the wavenumber selection is presented in detail in ref. \citep{hoylebook}.

Eq. \eqref{selected_wavenumber} suggests wavenumber, $q$ depends explicitly on the amplitude equation coefficients $\alpha$ and $\beta$. Eq. \eqref{crtick} also shows that $K_c$ expression contains $\alpha$ and $\beta$.  Now the coefficient, $\alpha$, contains the cross-diffusion terms. So by plotting wavenumbers against $\alpha$ in FIG. \ref{fig.wavenumbervsalpha}, we have shown the nature of implicit dependence of the wavenumber on the cross diffusion by varying both the cross-diffusion coefficients simultaneously while keeping all the other parameters, including the self-diffusion coefficients, fixed. Here the zero wavenumber point corresponds to the onset of the BF instability point. Therefore, it is possible to enter or leave the BF instability regime by tuning only the cross-diffusion coefficients of the system.  However, the continuous variation of wavenumber is only possible for infinite system consideration. Similarly, we have illustrated the variation of the complex amplitude's magnitude with $\alpha$ in the FIG. \ref{fig.magnitudevsalpha}. The magnitude sets the radius of the limit cycle in the system and from the FIG. 
\ref{fig.magnitudevsalpha}, it is evident that variation in the radius is possible by changing the cross-diffusion coefficient solely. As the wavenumber within the BF instability regime decreases towards the onset of BF instability point in FIG. \ref{fig.wavenumbervsalpha}, the magnitude increases gradually in \ref{fig.magnitudevsalpha}. A decline in the magnitude with a rise in the wavenumber is observed outside the instability regime.  
\section{\label{cds}Concentration Dynamics of the System}  
We can have both Hopf instability and traveling waves within the same parametric regime of the control parameter, $\nu$, depending on whether the selected wavenumber is zero or finite nonzero. The temporal pattern in a reaction-diffusion system can be traced at the critical wavenumber of Hopf instability from the corresponding amplitude. The evolution equation of the concentration representing the uniform oscillation near the onset of Hopf instability can be written by exploiting the amplitude equation formalism as 
\begin{equation}
{z_I}_{H}={z_I}_{0}+A_{H}U_{cH}\exp{(i f_{cH}t)}+C.C.,
\label{hwave}
\end{equation}   
with ${z_I}_{0}$ being the time-independent uniform base state for the extended direction and $A_{H}$ being the amplitude part within the oscillatory regime given by eq. \eqref{trialeq}. 

The evolution equation of the concentration,${z_I}_{TW}$ representing traveling waves near the onset of Hopf instability have expression similar to the Hopf instability. However, unlike the Hopf instability, the amplitude part $A_{H}$ holds the spatial variation in the case of traveling wave owing to non-zero wavenumber. Hence, the final general form of concentration dynamics within oscillatory regime is given by
\begin{widetext}
	\begin{gather}
		\begin{pmatrix}
			x\cr
			y
		\end{pmatrix}=
		\begin{pmatrix}
			x_0\cr
			y_0
		\end{pmatrix}
		+ \sqrt{\lambda-q^2}
		\begin{pmatrix}
			2\cos{(\omega_{0}t+f_{cH} t+qr)}+2\sin{(\omega_{0}t+f_{cH} t+qr)}\cr
			-2\cos{(\omega_{0}t+f_{cH} t+qr)}
		\end{pmatrix}.
	\label{maineq}
	\end{gather}
\end{widetext}

It is important to note here that the cross diffusion in a two-variable reaction-diffusion model can generate diffusion-driven Turing instability in the system\citep{GAMBINO2013, Chung2007}. However, we have not considered Turing instability in this report. The spatiotemporal chaos can also emerge for large system sizes in the presence of cross diffusion\citep{Berenstein2012}. This chaotic behavior is also out of the scope of this study. The concentration dynamics in eq. \eqref{maineq} only captures the dynamic features of Hopf instability and traveling waves within the parametric regime of interest. For the nonequilibrium thermodynamic study of the spatiotemporal pattern in the presence of cross diffusion due to overlapping of Turing and Hopf instability, one can consult ref. \citep{pkgg}.

\section{\label{sec:thermodynamics}Thermodynamics of nonlinear phenomenon of Chemical Reaction Network}
Nonequilibrium thermodynamic framework of nonlinear dynamic phenomena\citep{Rao2016NonequilibriumThermodynamics, Falasco2018InformationPatterns} at steady state can generate the system's energetics near the onset of Hopf instability for uniform oscillation and traveling waves pattern. The thermodynamic description of the pattern would reveal more about controlling the system's pattern and performance through diffusion coefficients around the BF instability and parametric phase-reversal dynamics. 
\subsection{\label{cle}Conservation Laws and Emergent Cycles}  
The stoichiometric matrix of the reversible Selkov model in eq. \eqref{revcrn} is
\begin{gather}
	S_{\rho}^{\sigma}=
	\bordermatrix{ ~ & R_{1} & R_{2}&R_{3}\cr
		S&1 &-1&0\cr
		P&0&1&-1 \cr
		A&-1&0&0 \cr
		B&0&0&1\cr} .\label{st}
\end{gather}
The left null vectors corresponding to left null space of the  stoichiometric matrix $S_{\rho}^{\sigma}$ are defined as the conservation laws\citep{Alberty2003ThermodynamicsReactions} and can be obtained from the expression 
\begin{equation}
\sum_{\sigma}{l_{\sigma}^{\lambda}S_{\rho}^{\sigma}}=0 \label{cons}
\end{equation} 
where $$\{l_{\sigma}^{\lambda}\}\in \mathbb{R}^{(\sigma-w )\times \sigma}, w=rank(S_{\rho}^{\sigma}).$$ For the stoichiometric matrix eq. \eqref{st} of the reversible Selkov model, the conservation law of the closed reaction network is,
\begin{gather}
	l_{\sigma}^{\lambda=1}=  
	\bordermatrix{~&X&Y&A&B\cr
		&1&1&1&1\cr}.\nonumber
\end{gather}
Components\citep{Alberty2003ThermodynamicsReactions}, conserved quantities of the chemical reaction network are defined as 
\begin{equation}
L_{\lambda}=\sum_{\sigma}{l_{\sigma}^{\lambda}}z_{\sigma}
\end{equation}
such that $\frac{d}{dt}\int dr L_{\lambda}=0$. Thus, component corresponding to the  conservation law is $L_1=s+p+a+b$. The conservation law of the system, $l_{\sigma}^{\lambda=1}$ is broken as the system is opened by chemostatting. Therefore,  corresponding component of the open system is no longer a global conserved quantity. The Right null space of the stoichiometric matrix, $S_{\rho}^{\sigma}c_{\sigma}^n$  represents internal cycle. However, this chemical reaction network has no internal cycle. The right null eigen vector corresponding to  null space of $S_{\rho}^{I}$ is defined as the emergent cycle\citep{Polettini2014IrreversibleLaws}. States of intermediate species remain unchanged over a complete emergent cycle but chmeostatted species are exchanged between system and chemostats. The total number of chemostatted species is equal to the sum of the number of broken conservation laws and the number of emergent cycles in open chemical reaction network\citep{Polettini2014IrreversibleLaws, Rao2016NonequilibriumThermodynamics}. Therefore, the reversible Selkov model has one independent emergent cycle 
\begin{gather}  
	c_{1}^{}=  
	\bordermatrix{~\cr
		1&1\cr
		2&1\cr
		3&1}.\nonumber
\end{gather} 

\subsection{\label{epr}Entropy Production Rate}
The forward or reverse flux corresponding to an elementary reaction can be expressed in accord with mass action law by, $j_{\pm \rho}=k_{\pm\rho}\prod_{\sigma}z^{v_{\pm\rho}^{\sigma}}_{\sigma}$ with $'+'$ and $'-'$ label forward and backward reaction, respectively and  $v_{\pm \rho}^{\sigma}$ denotes the number of molecules of a particular species $'\sigma'$. Thus the net flux  will be   $j_{\rho}=j_{+\rho}-j_{-\rho}$. The thermodynamic driving forces of reaction known as reaction affinities\citep{Prigogine1954ChemicalDefay.} is given by
$f_{\rho}=-\sum_{\sigma}{S_{\rho}^{\sigma}\mu_{\sigma}}$
where $S_{\rho}^{\sigma}=v_{{-}\rho}^{\sigma}-v_{{+}\rho}^{\sigma}$ \text{ is the stoichiometric coefficient of species and}  $\mu_{\sigma}=\mu_{\sigma}^o+\ln{\frac{z_{\sigma}}{z_0}}$ is the chemical potential with solvent concentration $z_0$ and standard-state chemical potential $\mu_{\sigma}^o$. The nonequilibrium concentrations of intermediate species solely attribute to the global nonequilibrium state of the chemical reaction network. Hence implementing the equilibrium form of thermodynamic variables in this nonequilibrium framework can be justified under the assumption that the nonequilibrium system is kept at local thermal equilibrium at a temperature set by the solvent in a dilute solution. The system is fixed at constant absolute temperature $T$ by the solvent, and  $RT$ is considered as unity here. From the expression of affinities one can further write another form in terms of the reaction fluxes of the chemical steps as $f_{\rho}= \ln{\frac{j_{+\rho}}{j_{-\rho}}}$. Therefore, the entropy production rate(EPR) due to the chemical reaction can be expressed  using flux-force relation as

\begin{equation}
\frac{d\Sigma_{R}}{dt}=\frac{1}{T}\int  dr \sum_{\rho} (j_{+\rho}-j_{-\rho}) \ln{\frac{j_{+\rho}}{j_{-\rho}}},
\label{eprr}
\end{equation}

Now, Considering diffusive flux and affinity, the entropy production rate due to diffusion can be expressed as
 
\begin{eqnarray}
\frac{d\Sigma_{D}}{dt}=\int dr \Big[ D_{11}\frac{{\parallel{\frac{\partial x}{\partial r} }\parallel}^2}{x}+D_{22}\frac{{\parallel{\frac{\partial y}{\partial r} }\parallel}^2}{y}\nonumber\\ +D_{12}\frac{{\parallel{\frac{\partial y}{\partial r}\parallel} {\parallel\frac{\partial x}{\partial r} }\parallel}}{x}+
D_{21}\frac{{\parallel{\frac{\partial x}{\partial r}\parallel} {\parallel\frac{\partial x}{\partial r} }\parallel}}{y}\Big] \label{eprdd}.
\end{eqnarray}
The last two terms on the right in eq. \eqref{eprdd} represent the contribution of the cross diffusion of the intermediate species.

Total EPR comprises of homogeneous part EPR, reaction EPR and diffusion EPR.
Under the second law of thermodynamics, the total EPR will always be positive.

\subsection{\label{sgfe}Semigrand Gibbs Free Energy}
We need to take the open system's true thermodynamic potentials to analyze the chemical reaction network's energetics at the nonequilibrium regime. The nonequilibrium Gibbs free energy of a  reaction network can be expressed in terms of the chemical potential as\citep{Fermi1956Thermodynamics} 
\begin{equation}
G=G_{0}+ \sum_{\sigma \neq 0}{(z_{\sigma}\mu_{\sigma}-z_{\sigma})}\label{neg}
\end{equation}
with $G_{0}=z_{0}\mu_{0}^o$. However, to capture the energetics of the open system properly one needs to define  the semigrand Gibbs
free energy\citep{Falasco2018InformationPatterns} of the system analogous to the grand potential of  the grand canonical ensemble. The semigrand Gibbs free energy of an open system can be acquired  by operating a Legendre transformation\citep{Rao2016NonequilibriumThermodynamics} on the nonequilibrium Gibbs free energy
\begin{equation}
\mathcal{G}=G-\sum_{\lambda_b}{\mu_{\lambda_b}M_{\lambda_b}}. 
\label{smgg}
\end{equation}
where $M_{\lambda_b}=\sum_{C_b}l_{C_b}^{{\lambda_b}^{-1}} L_{\lambda_b}$ represents moieties exchanged between chemostats and system. Now the affinities acting along emergent cycles of the system obeys following equation
\begin{equation}
\mu_{\epsilon}=c_{\epsilon}\ln{\frac{k_{\rho}}{k_{-\rho}}}z_{c}^{-S_{\rho}^c},
\end{equation} 
where $z_{c}$ and $S_{\rho}^c$ are respectively the concentrations and stoichiometric elements corresponding to the chemostatted species. 
\section{\label{redi}Results and Discussion} 
\begin{figure*}[htbp!]
	\centering 
	\subfigure[\label{fig.hopfphase}]{\includegraphics[width=0.33\textwidth]{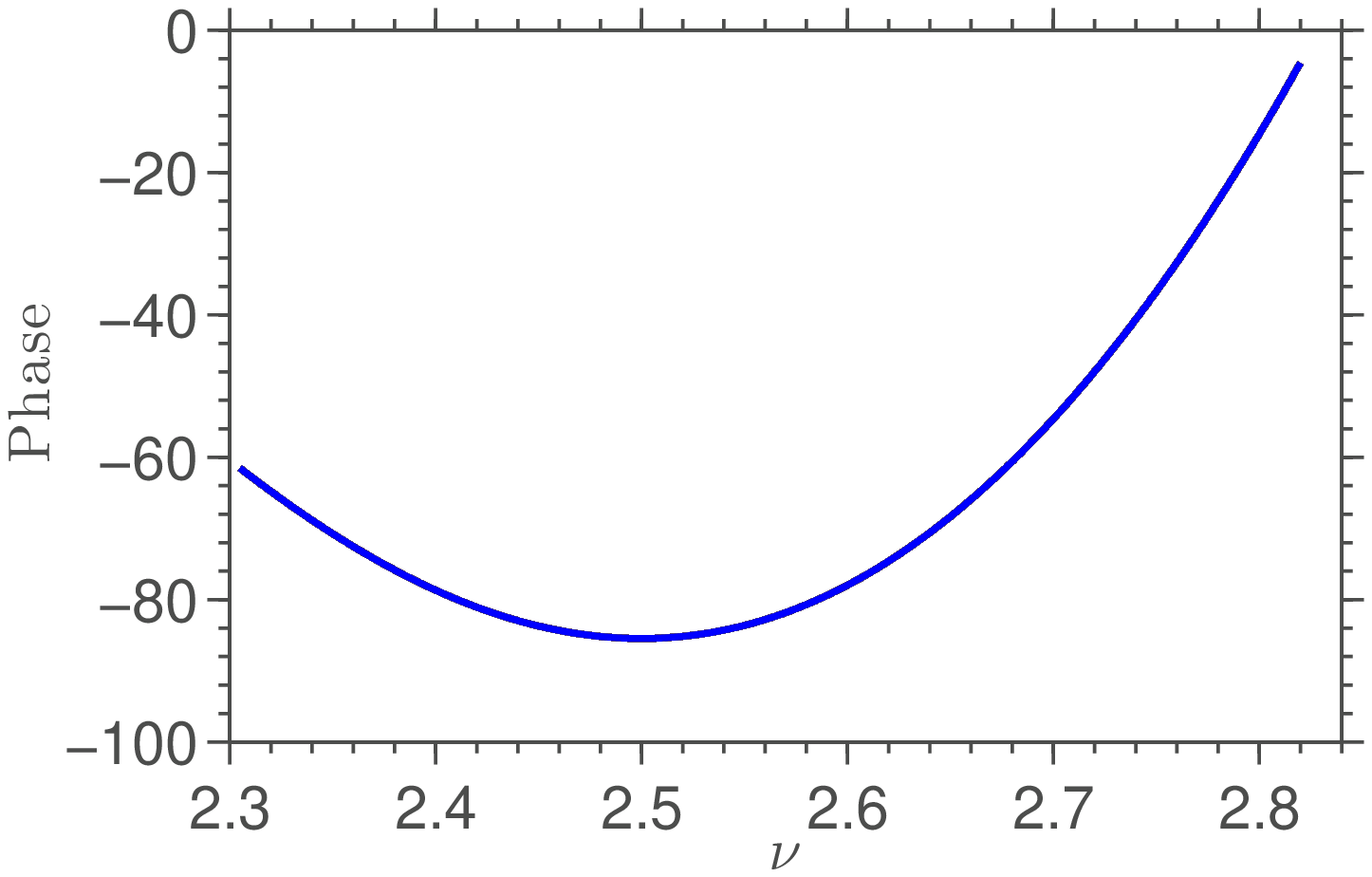}}\hfill
	\subfigure[\label{fig.wavephase}]{\includegraphics[width=0.33\textwidth]{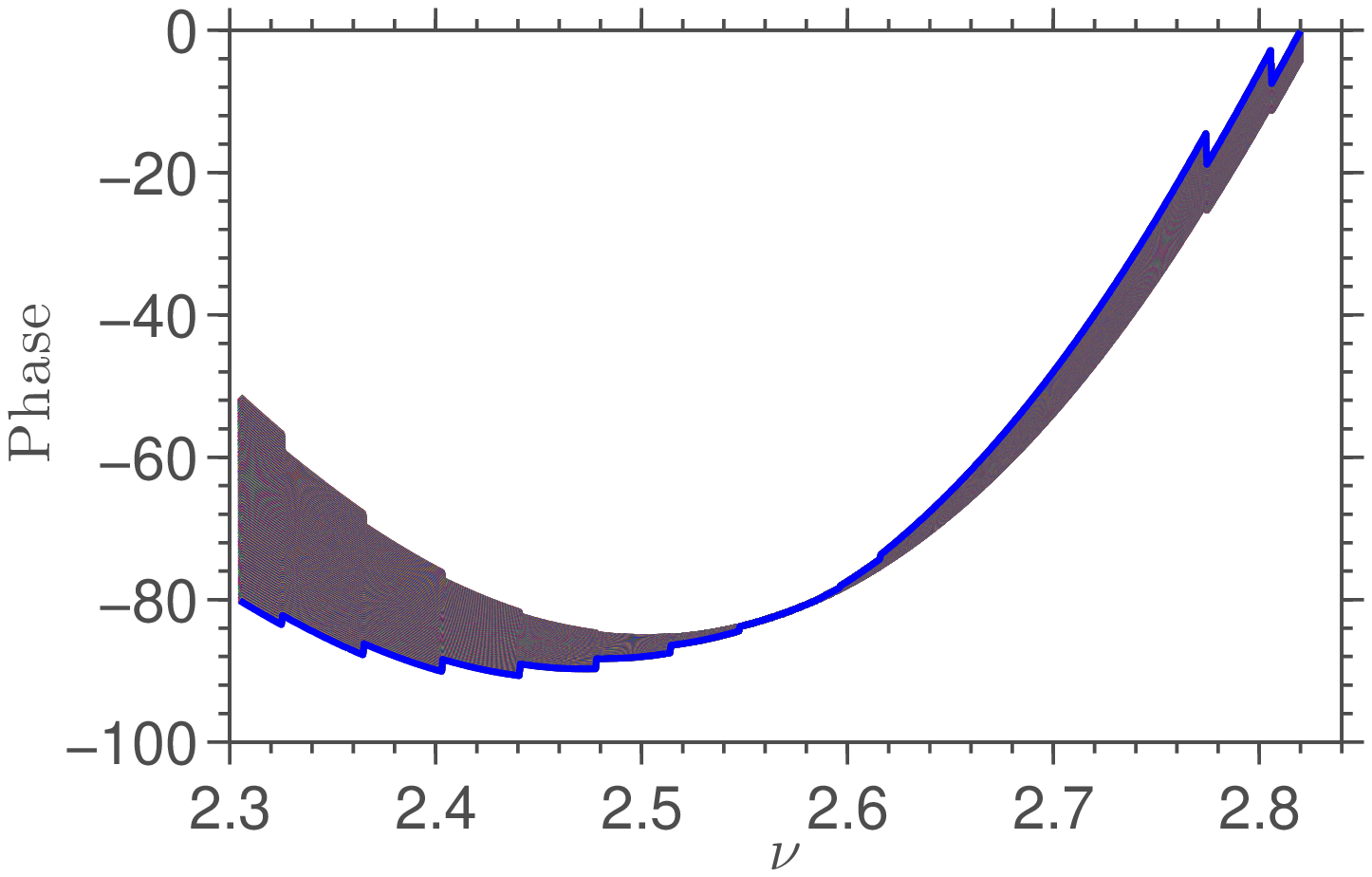}}
\hfill
\subfigure[\label{fig.wavephasein}]	
{\includegraphics[width=0.33\textwidth]{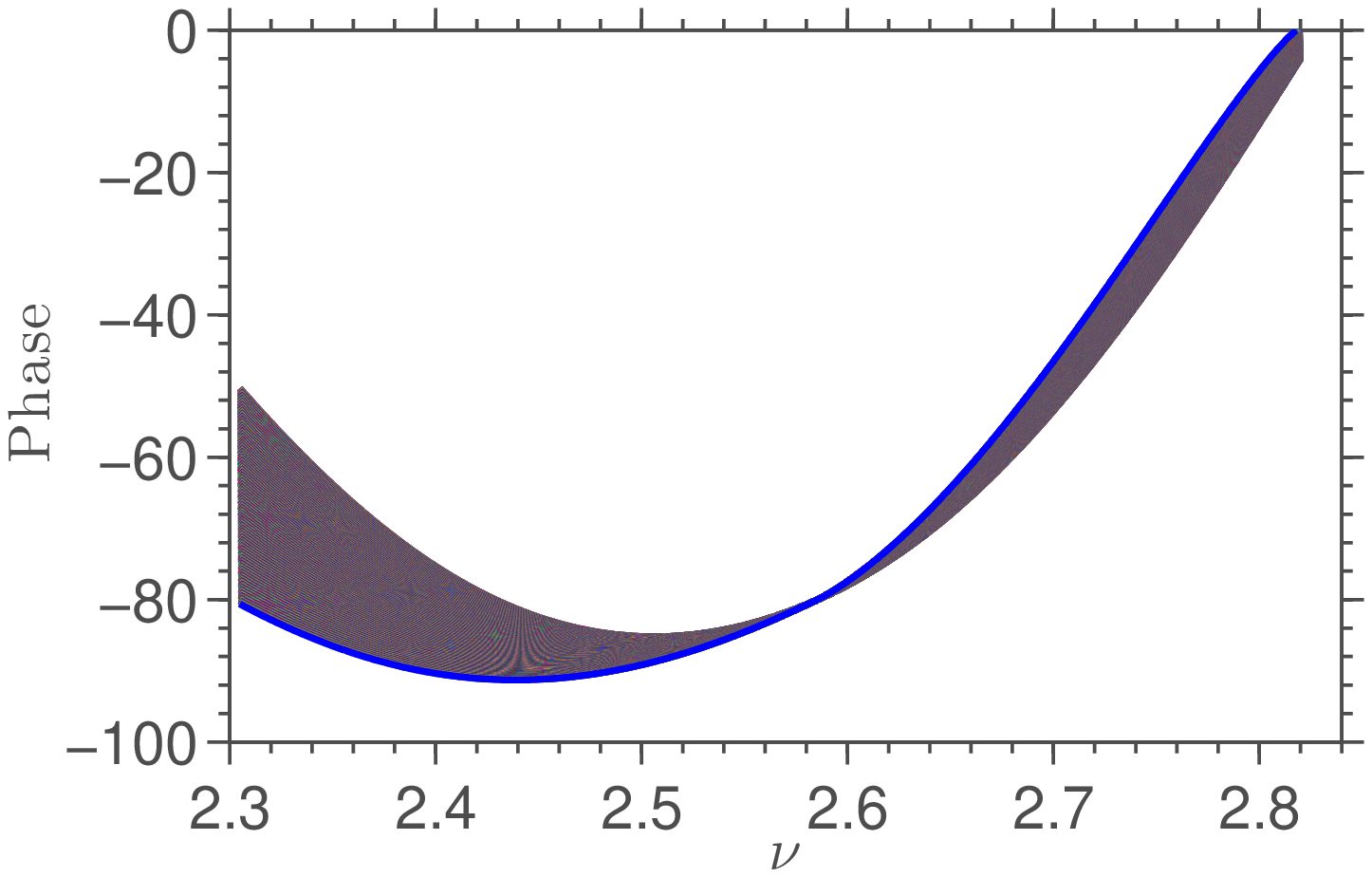}}\hfill	
\subfigure[\label{fig.namp}]	
{\includegraphics[width=0.33\textwidth]{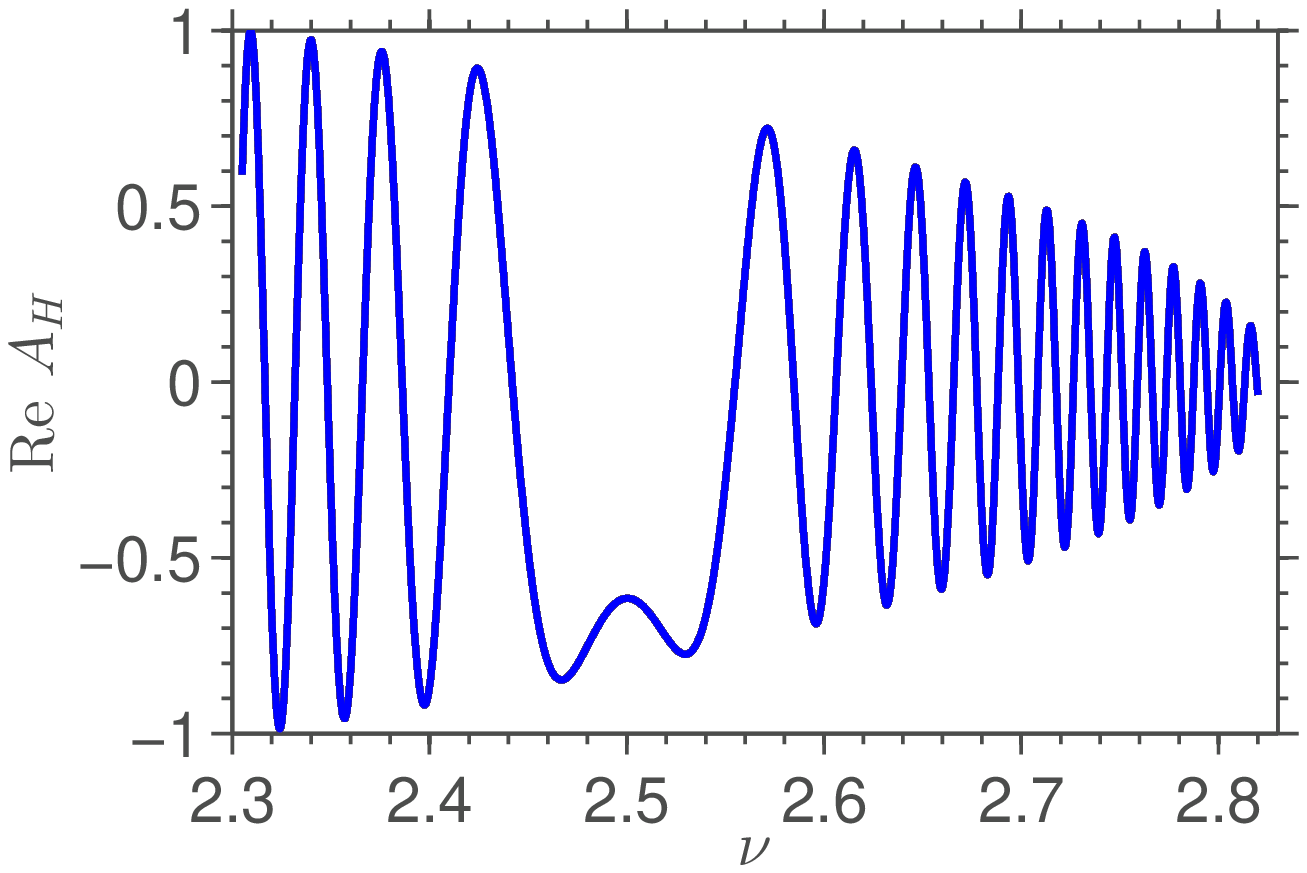}}\hfill
	\subfigure[\label{fig.wavnamp}]{\includegraphics[width=0.33\textwidth]{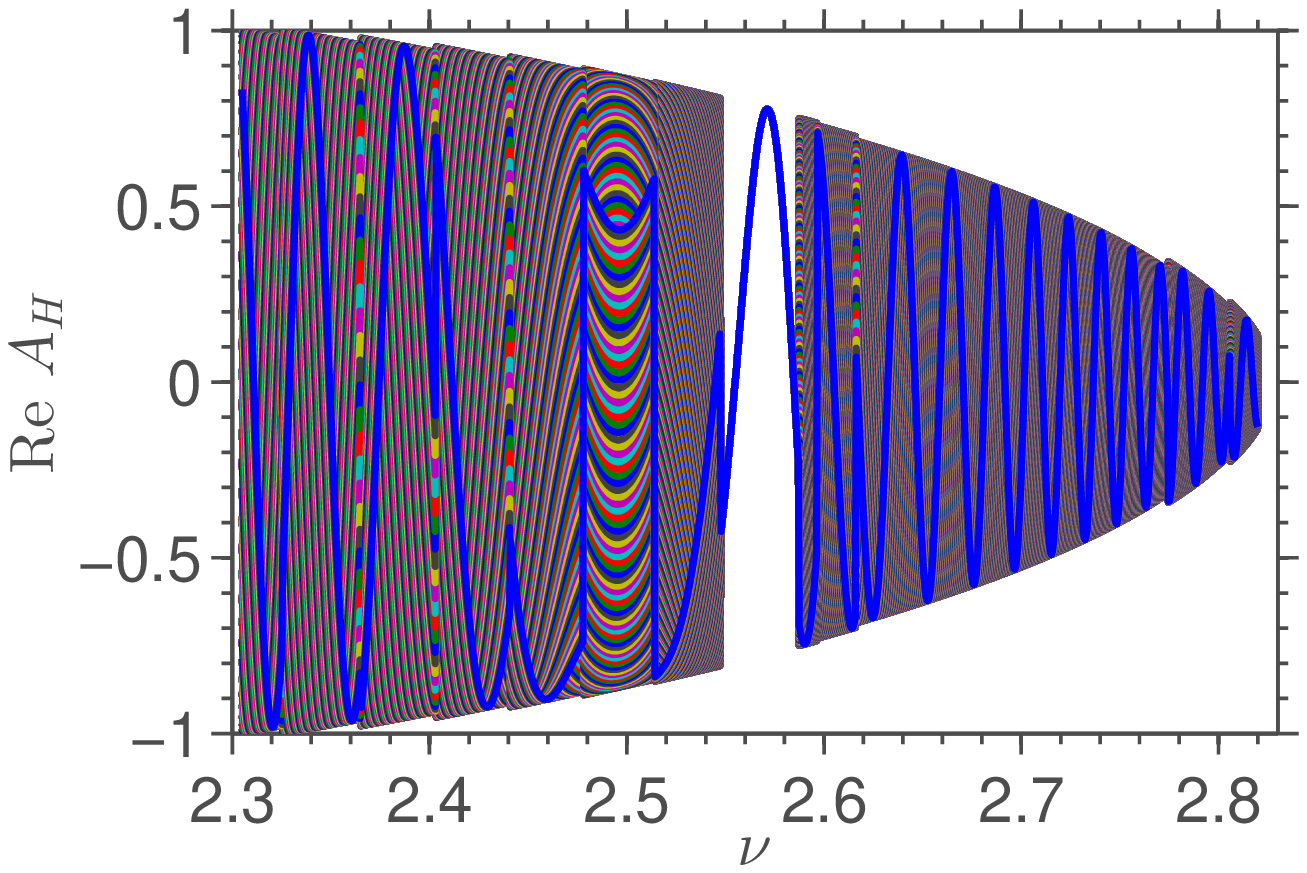}}\hfill	
\subfigure[\label{fig.wavnampin}]{\includegraphics[width=0.33\textwidth]{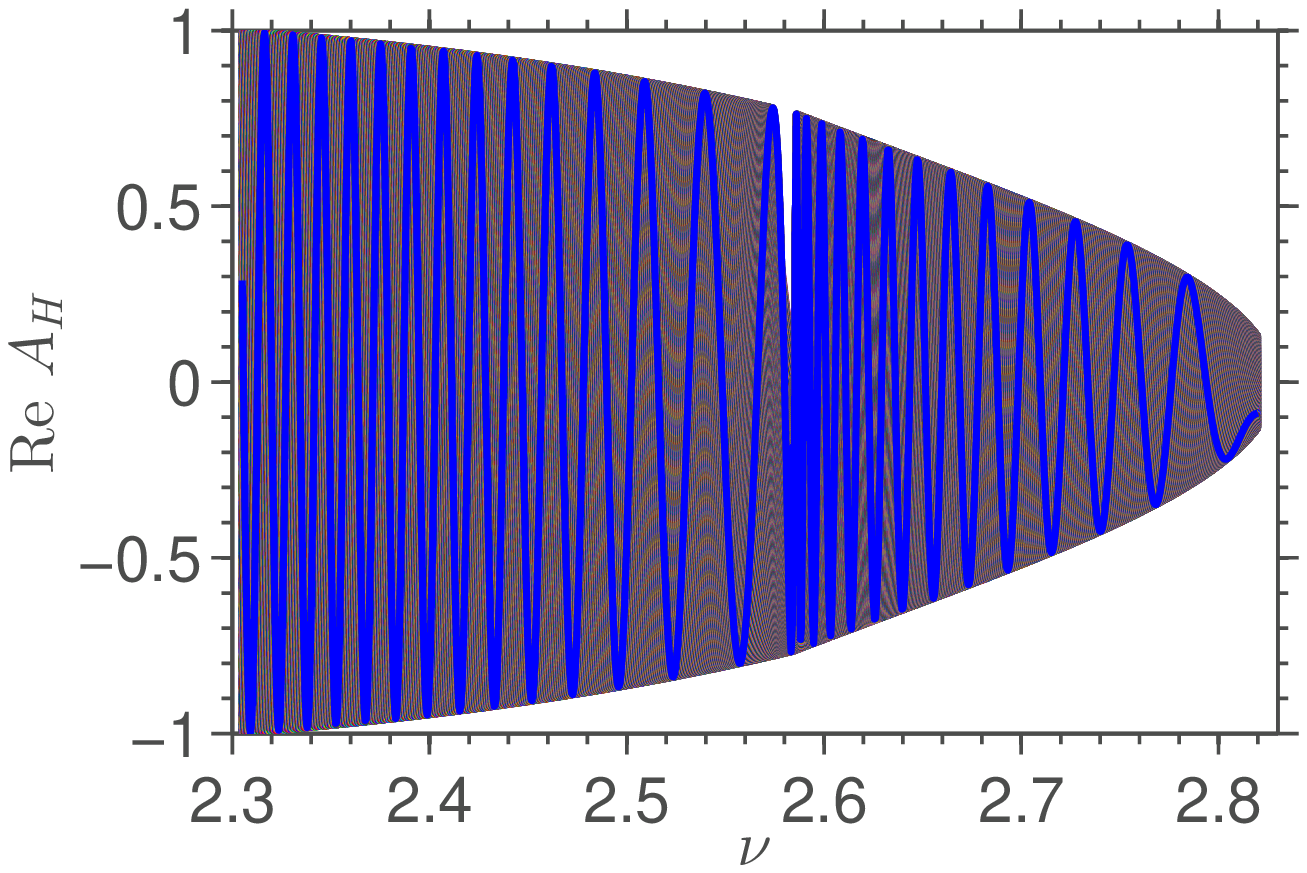}}\hfill
\caption{ \label{ampphasedynamics} The system's phase as a function of the control parameter, $nu$, is illustrated in FIG. \ref{fig.hopfphase} and \ref{fig.wavephase} for Hopf instability and traveling waves, respectively for finite domain of length $l=200$. The phase change in the limit of infinite size is illustrated in FIG. \ref{fig.wavephasein}.  The normalized real part of Hopf amplitude denoted by `Re $A_H$' comprised of magnitude and phase is obtained analytically as a function of control parameter $\nu$  at the time, $t=400$ in FIG. \ref{fig.namp}. Corresponding normalized real part of amplitude fields for traveling waves for discrete and continuous wavenumber are shown in FIG. \ref{fig.wavnamp} and \ref{fig.wavnampin}, respectively. These amplitude figures will provide a better understanding of the local concentration profile in the 1D Selkov model in the parameter space of Hopf instability and traveling waves. Here, diffusion  coefficients are: $D_{11}=D_{22}=0.0005
 1;D_{12}=-0.0002;D_{21}=0.0002$ and the value of the parameter $\omega$ is set as 2.}
\end{figure*}

\begin{figure*}[htbp!]
	\centering 
	\subfigure[\label{fig.x}]{\includegraphics[width=0.33\textwidth]{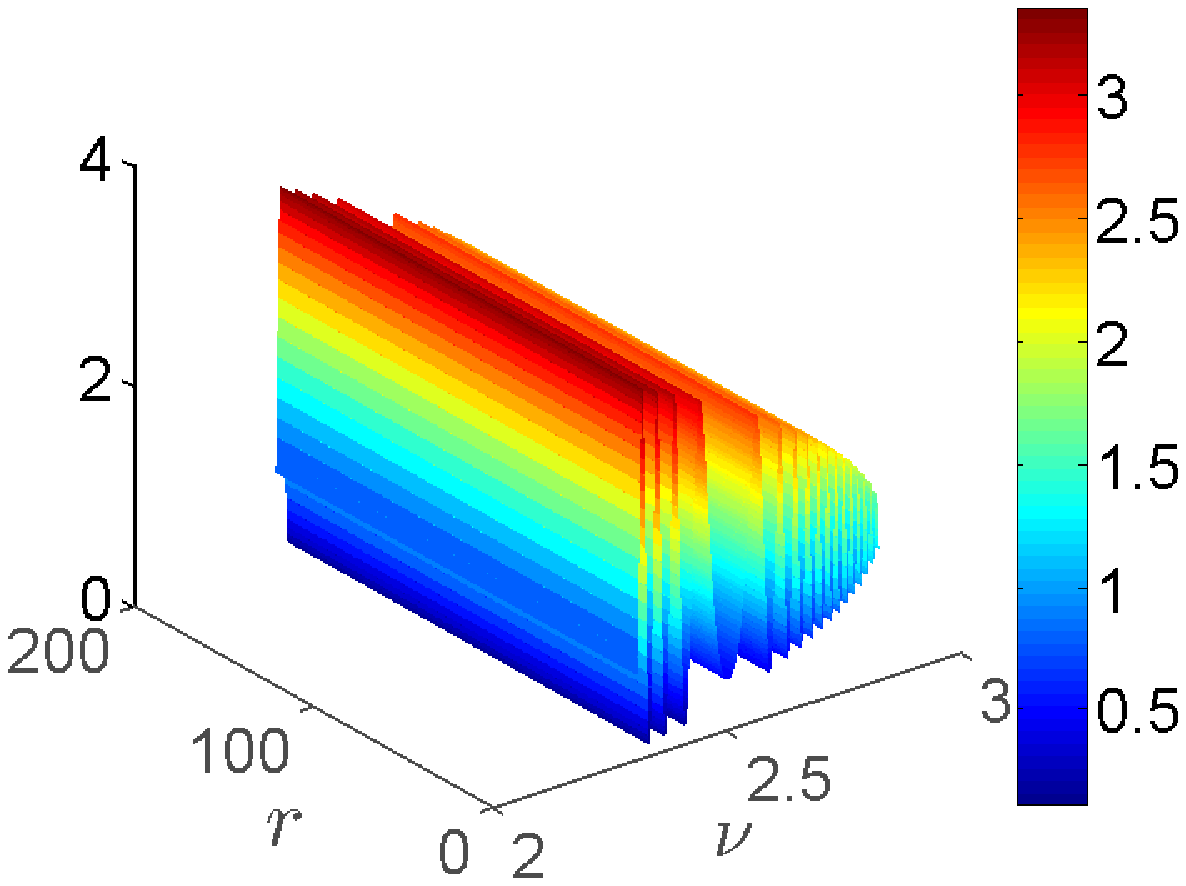}}\hfill
	\subfigure[\label{fig.wavx}]{\includegraphics[width=0.33\textwidth]{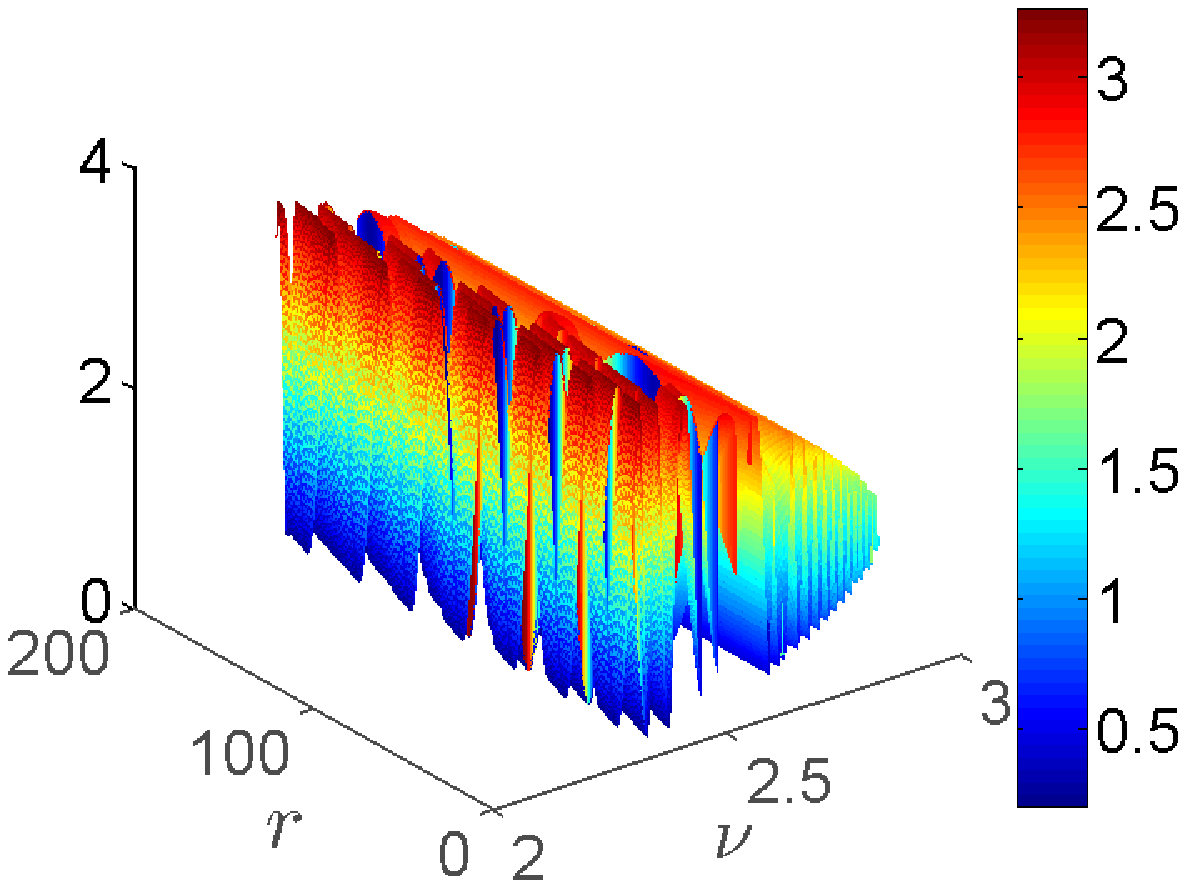}}\hfill
	\subfigure[\label{fig.wavxin}]{\includegraphics[width=0.33\textwidth]{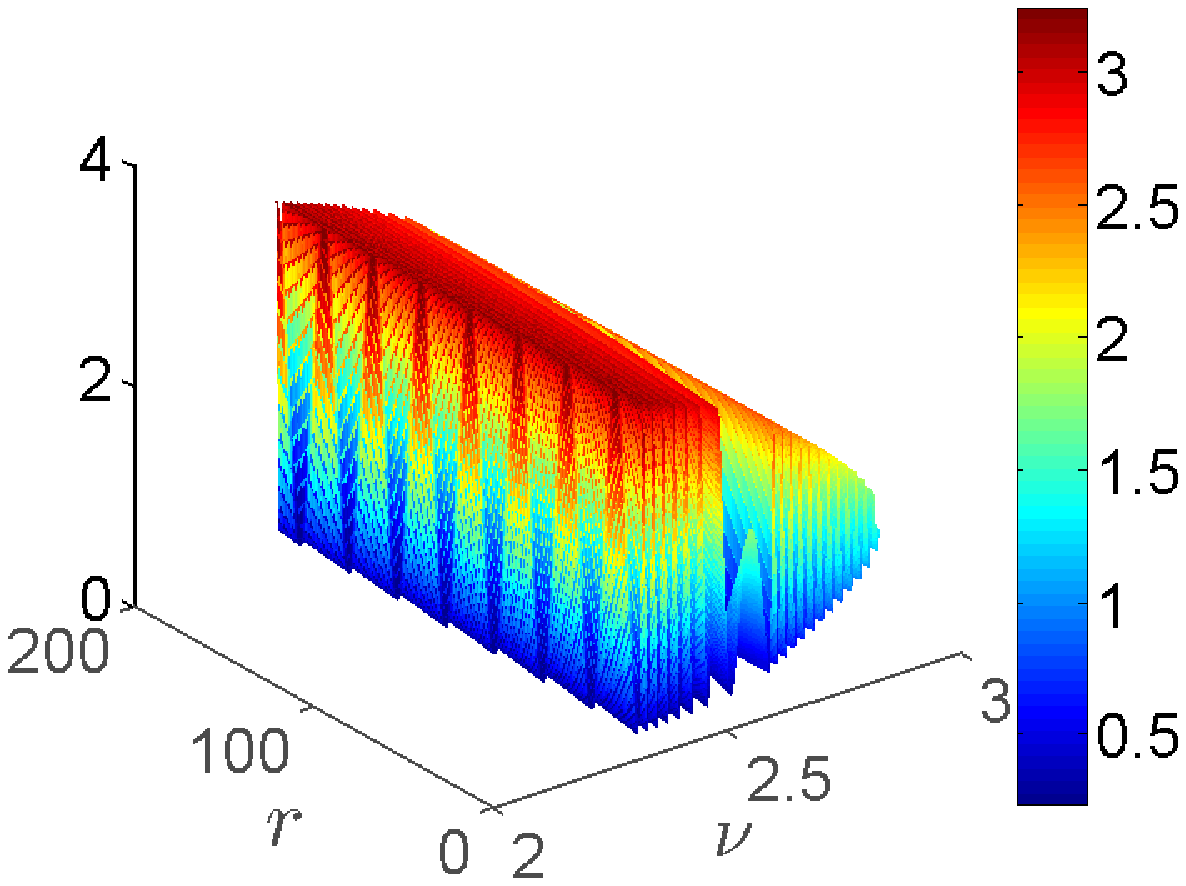}}\hfill
	\subfigure[\label{fig.ximage}]{\includegraphics[width=0.33\textwidth]{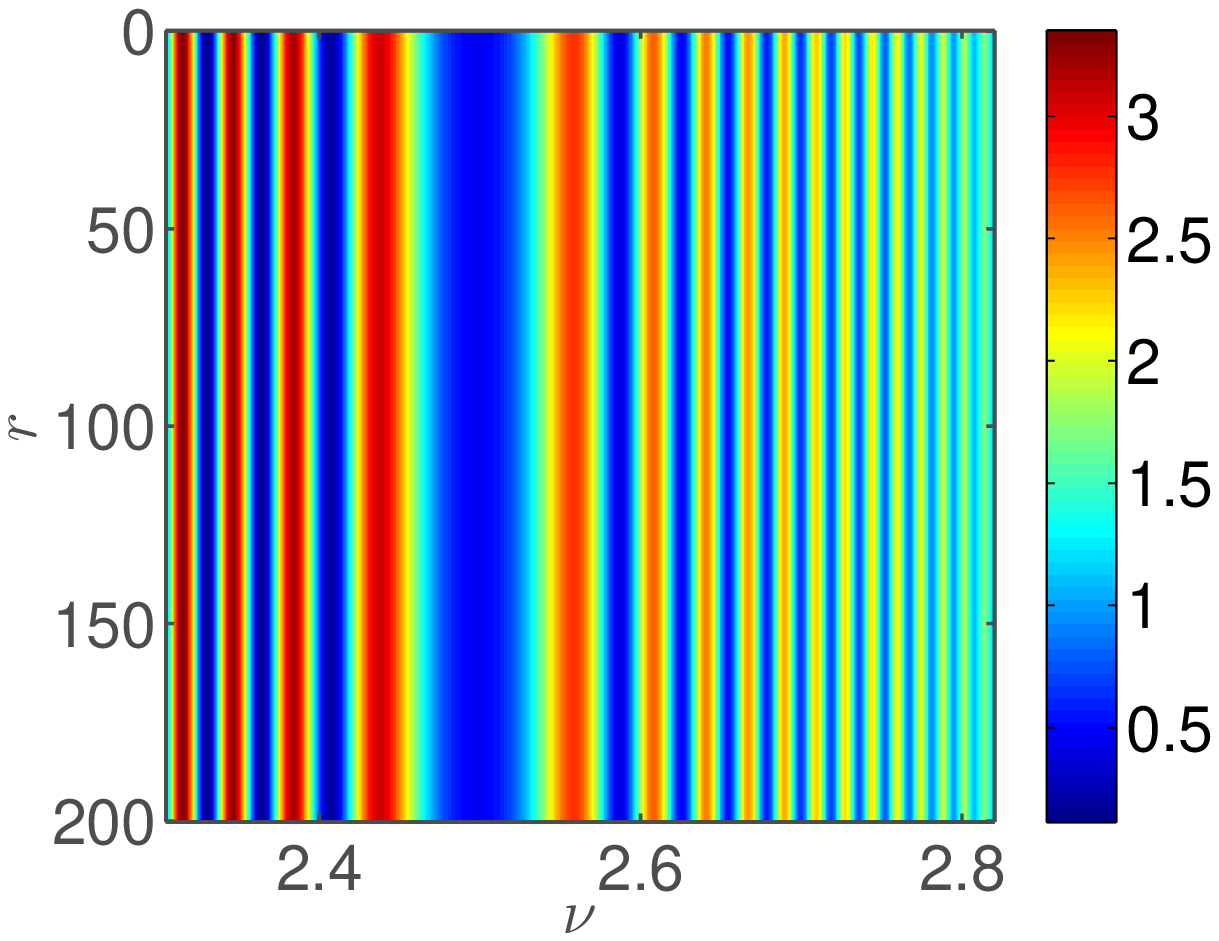}}\hfill
\subfigure[\label{fig.wavximage}]{\includegraphics[width=0.33\textwidth]{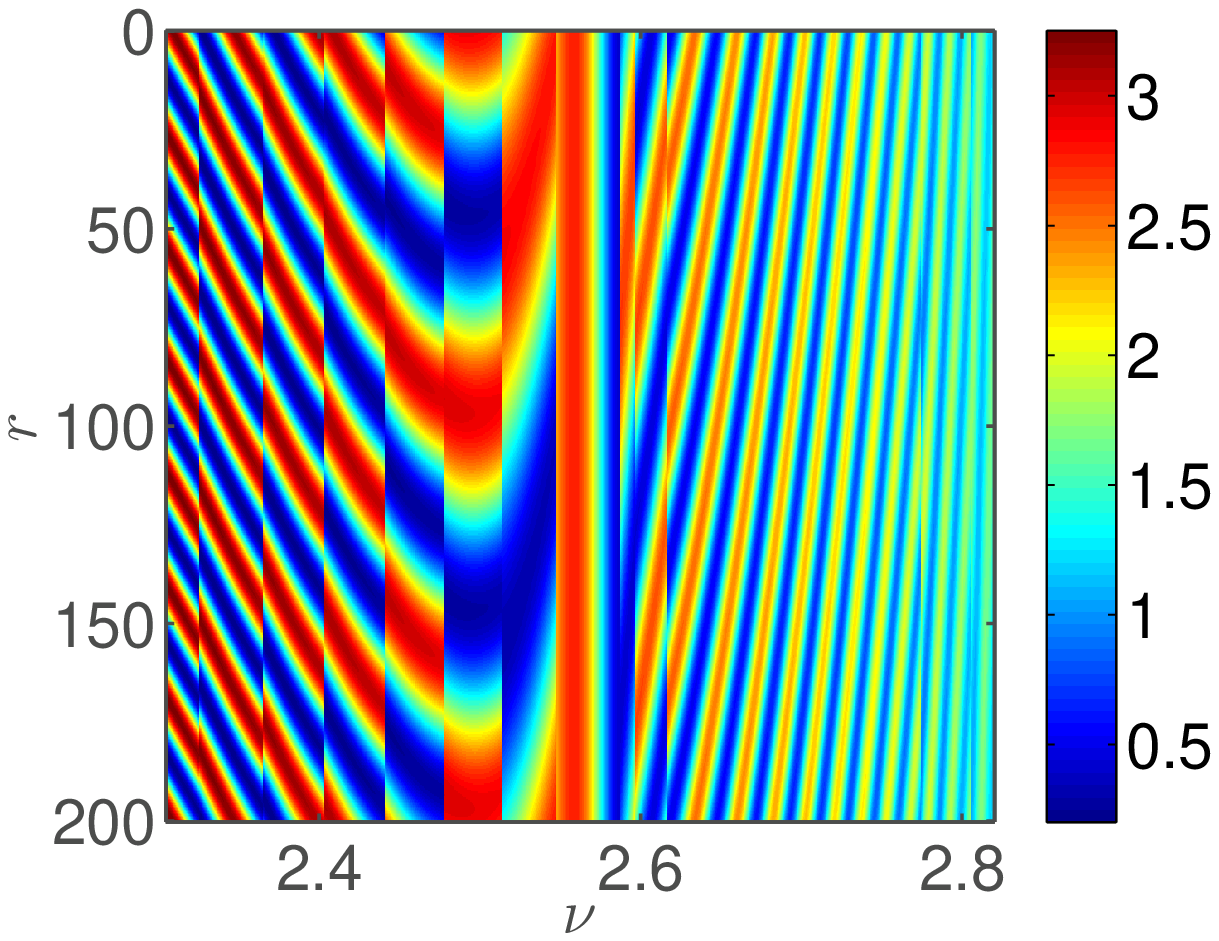}}\hfill	
\subfigure[\label{fig.wavximagein}]{\includegraphics[width=0.33\textwidth]{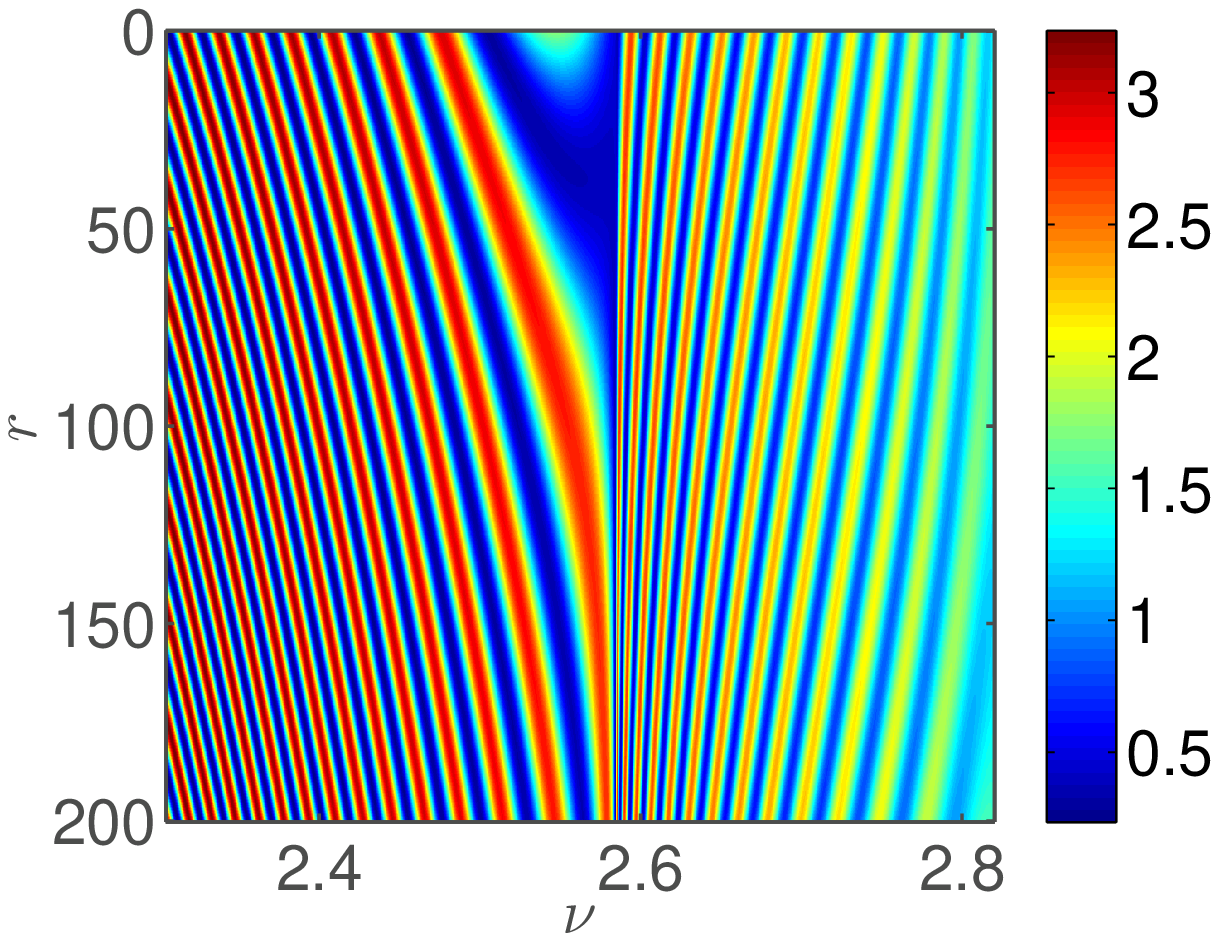}}
\caption {In FIG. \ref{fig.x}, the 3d concentration field of $X$ for the Hopf instability in the Selkov model of length $l=200$ as a function of externally controlled parameter, $\nu$ is presented (plot of $Y$ is similar) at time, $t=400$ and temperature, $T=300K$ for a fixed value of parameter $\omega=2$. The 'jet' colormap is used to show contrast in concentration field. The FIG. \ref{fig.ximage} illustrates the corresponding image of the concentration field of $X$. The 3d concentration field of $X$ for the same system in the case of travelling waves with finite system consideration is illustrated in FIG. \ref{fig.wavx} and corresponding image is shown in FIG. \ref{fig.wavximage}. The extended spatial dimension is considered along the vertical axis. The 3d concentration field of $X$ and corresponding image for traveling waves with continuous family of wavenumbers in the limit of infinite size are demonstrated in FIG. \ref{fig.wavxin} and \ref{fig.wavximagein}, respectively.
  Diffusion  coefficients are: $D_{11}=D_{22}=0.00051;D_{12}=-0.0002;D_{21}=0.0002$ and all the reactions are weakly reversible i.e. $K_{-\rho}=10 ^{-4}$.}
	\label{xrt}
\end{figure*}
\begin{figure*}[htbp!]
	\centering 
	\subfigure[\label{fig.xline}]{\includegraphics[width=0.46\textwidth]{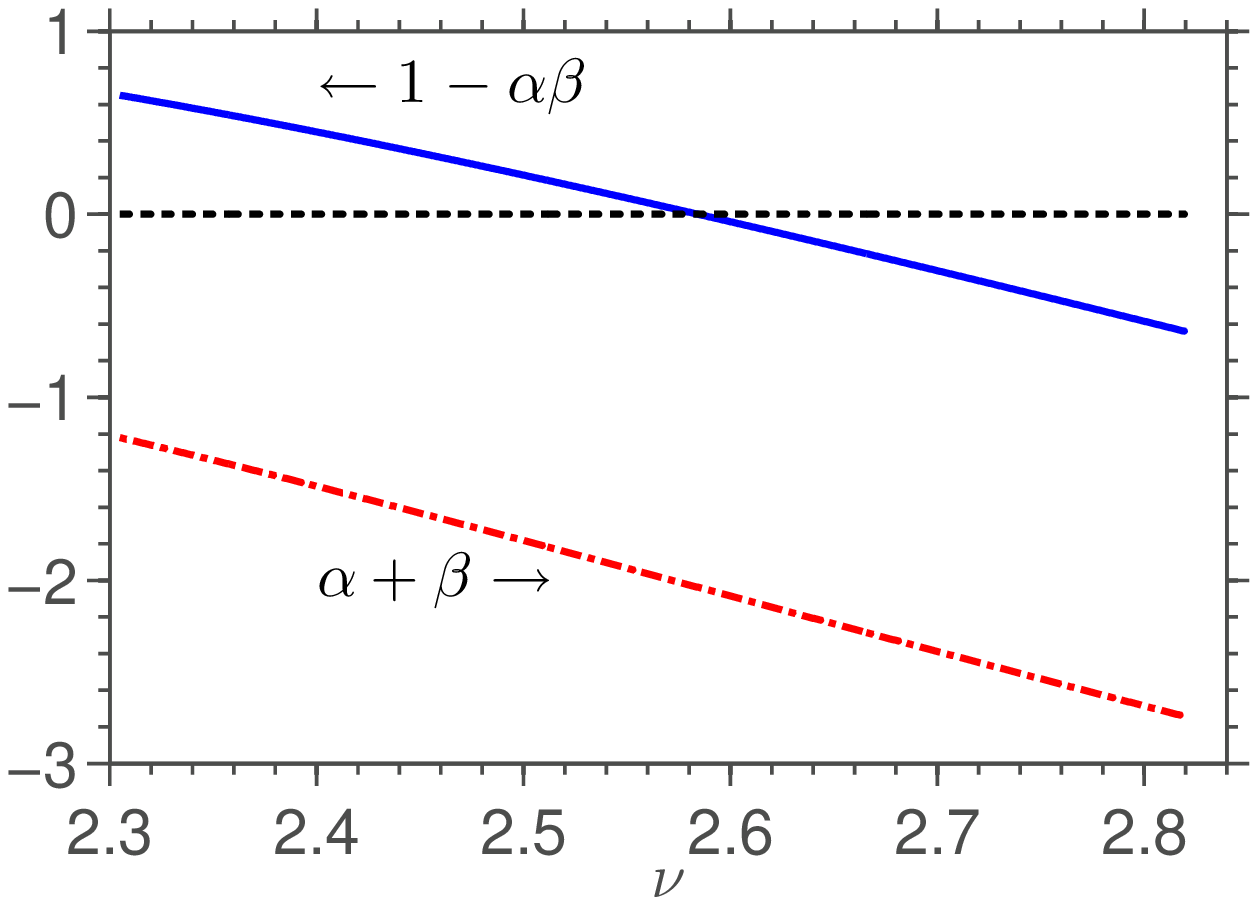}}\hfill
	\subfigure[\label{fig.limit}]{\includegraphics[width=0.46\textwidth]{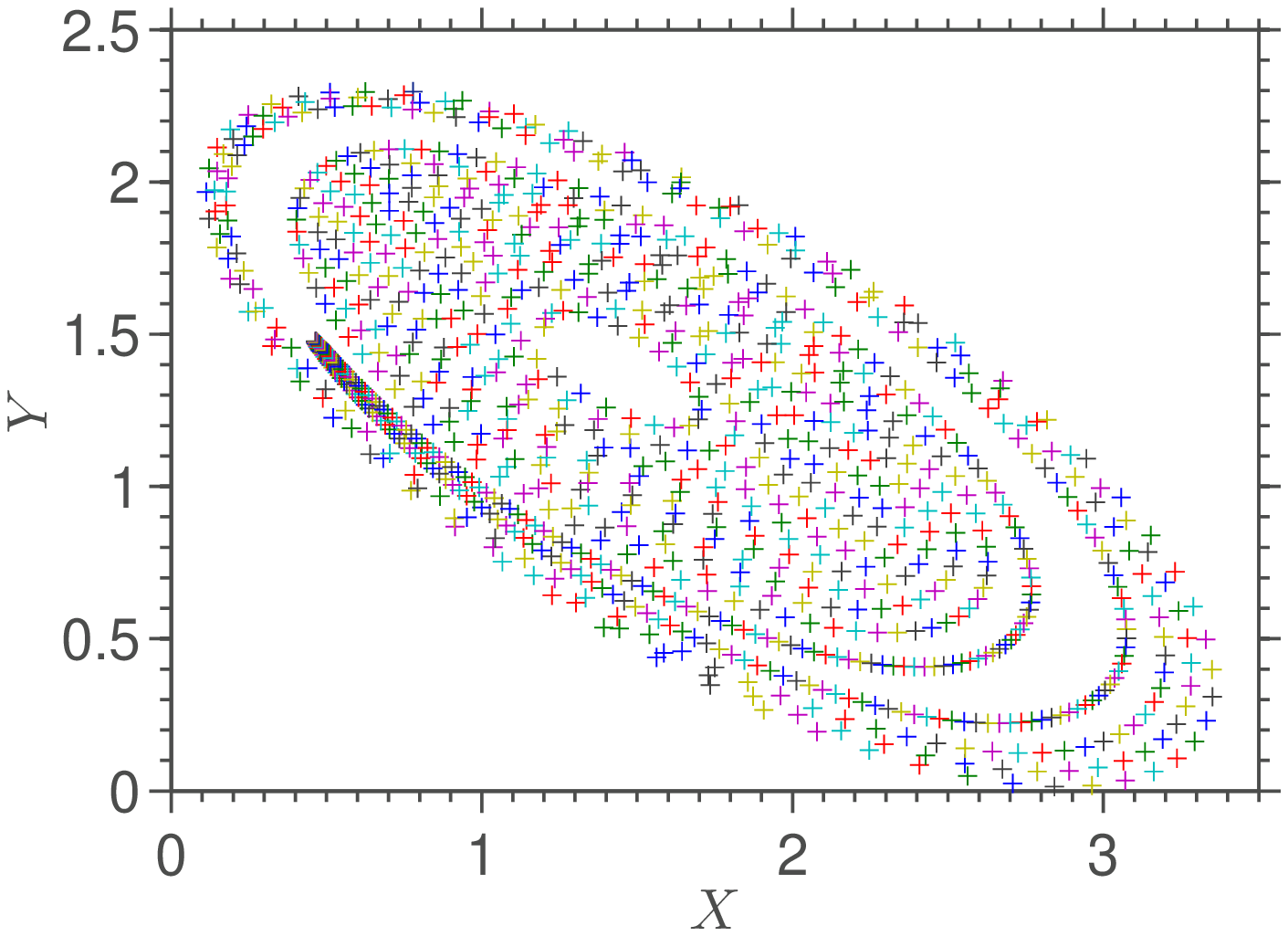}}\hfill
	\label{phasechange}
	\caption{The Benjamin-Feir(BF) instability and phase reversal defining conditions corresponding to amplitude equation are shown in FIG. \ref{fig.xline}. FIG. \ref{fig.limit} illustrates a plot similar to a phase portrait in the case of Hopf instability. Here $X$ and $Y$ concentrations dynamics are obtained by varying $\nu$ but for a fixed time $t=400$. The red dotted line in FIG. \ref{fig.xline} corresponds to the $\alpha + \beta$ and the solid blue line represents $1-\alpha\beta$. As $1-\alpha\beta$ line crosses the zero line(the dotted black line), we enter the BF instability regime. Diffusion  coefficients are: $D_{11}=D_{22}=0.00051;D_{12}=-0.0002;D_{21}=0.0002$ and all the reactions are weakly reversible i.e. $K_{-\rho}=10 ^{-4}$.}
\end{figure*}

\begin{figure}
	\centering
\includegraphics[width=\linewidth]{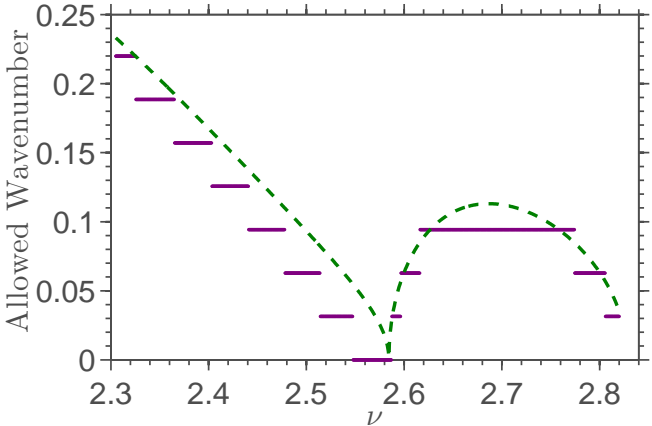}
 \caption{\label{fig.wavenumber} The wavenumber near the onset of Hopf instability. For the finite system of length, $l=200$, we have shown discrete allowed values of wavenumber below the continuous wavenumber. The continuous wavenumber in the limit of infinite size, is illustrated by dashed line. Here, diffusion  coefficients are: $D_{11}=D_{22}=0.00051;D_{12}=-0.0002;D_{21}=0.0002$ and the value of the parameter $\omega$ is set as 2.}
\end{figure}

\begin{figure*}[htp!]
	\centering 
	\subfigure[\label{fig.totalepr}]{\includegraphics[width=0.33\linewidth]{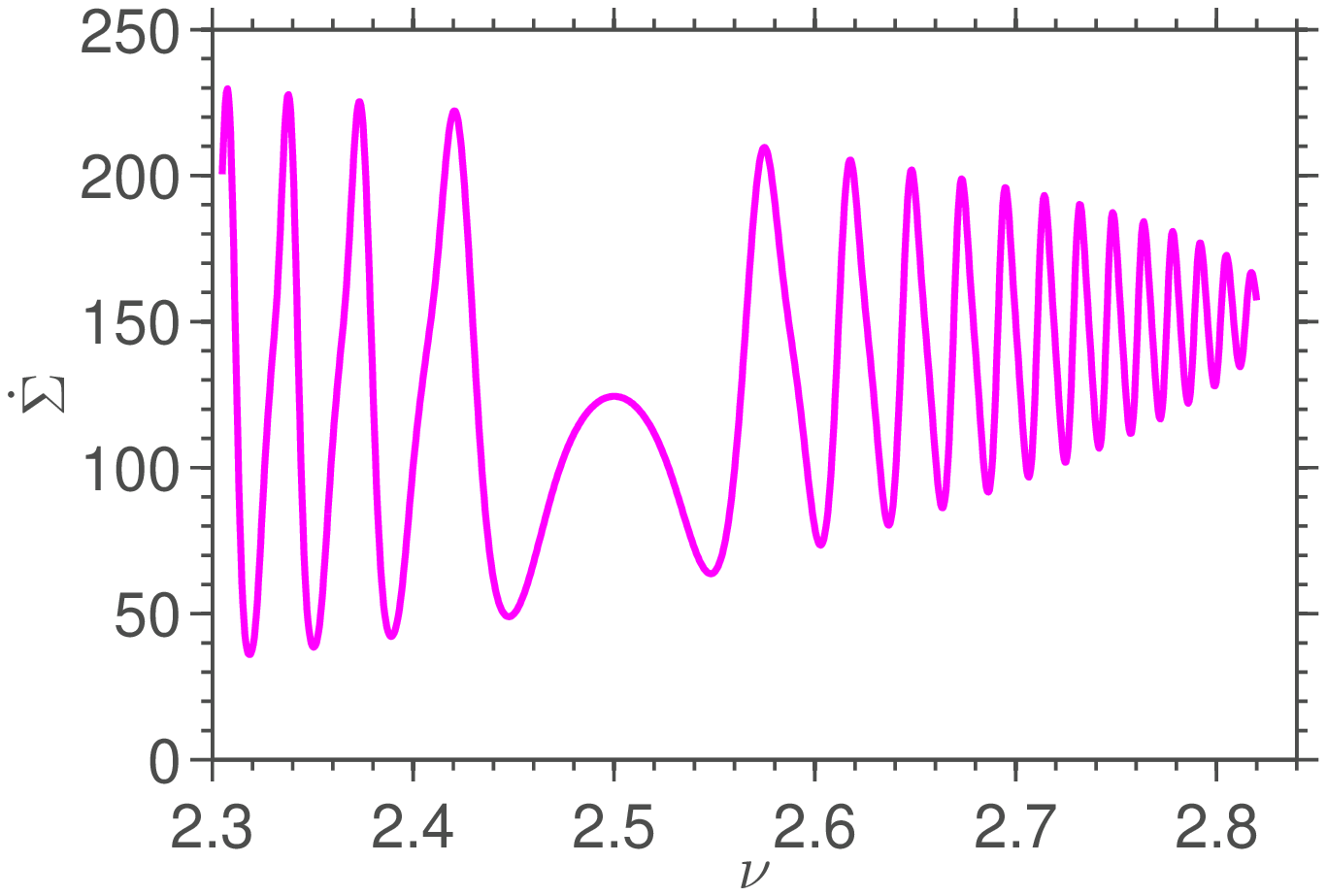}}\hfill
	\subfigure[\label{fig.wavtotalepr}]{\includegraphics[width=0.33\linewidth]{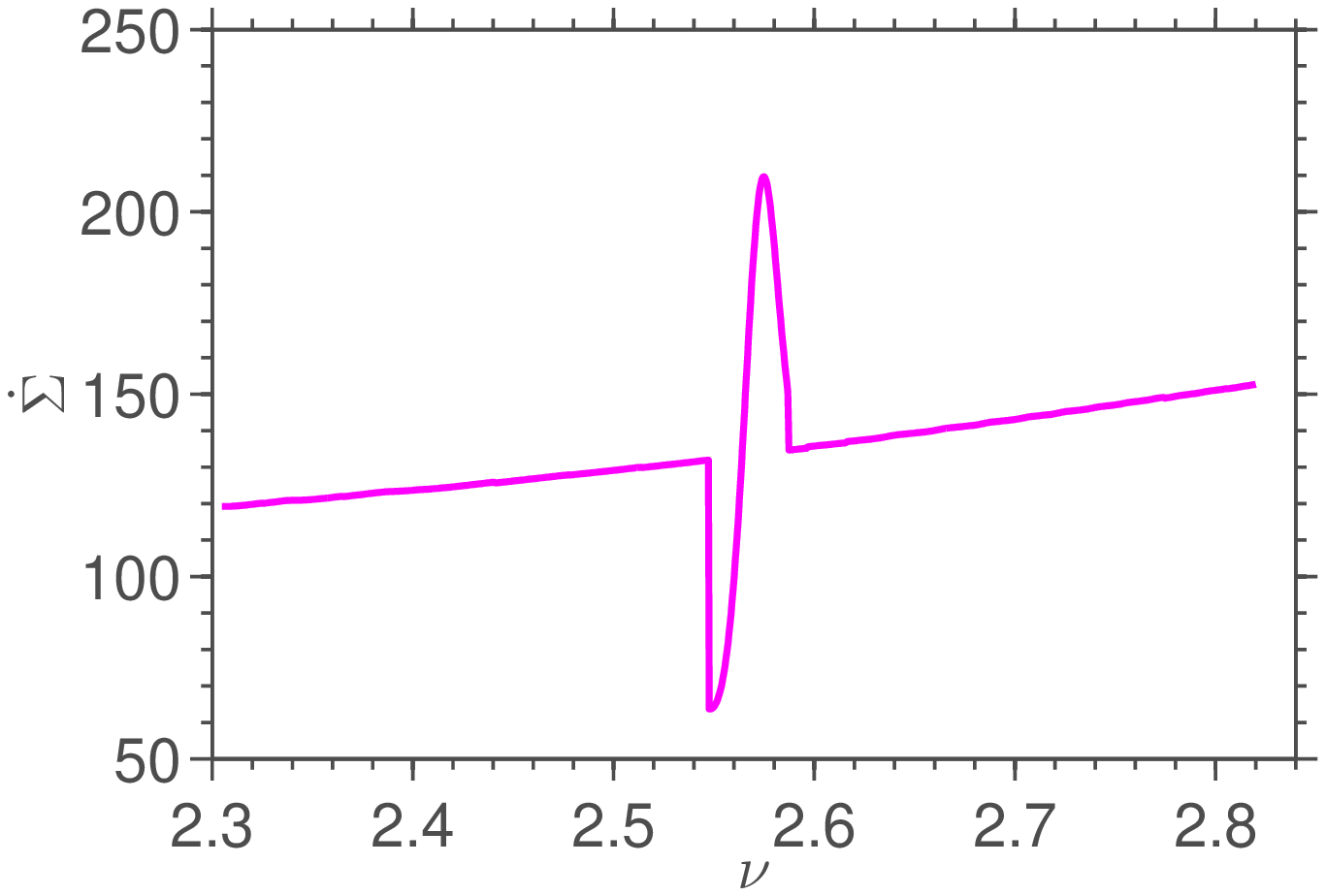}}\hfill
	\subfigure[\label{fig.wavtotaleprin}]{\includegraphics[width=0.33\linewidth]{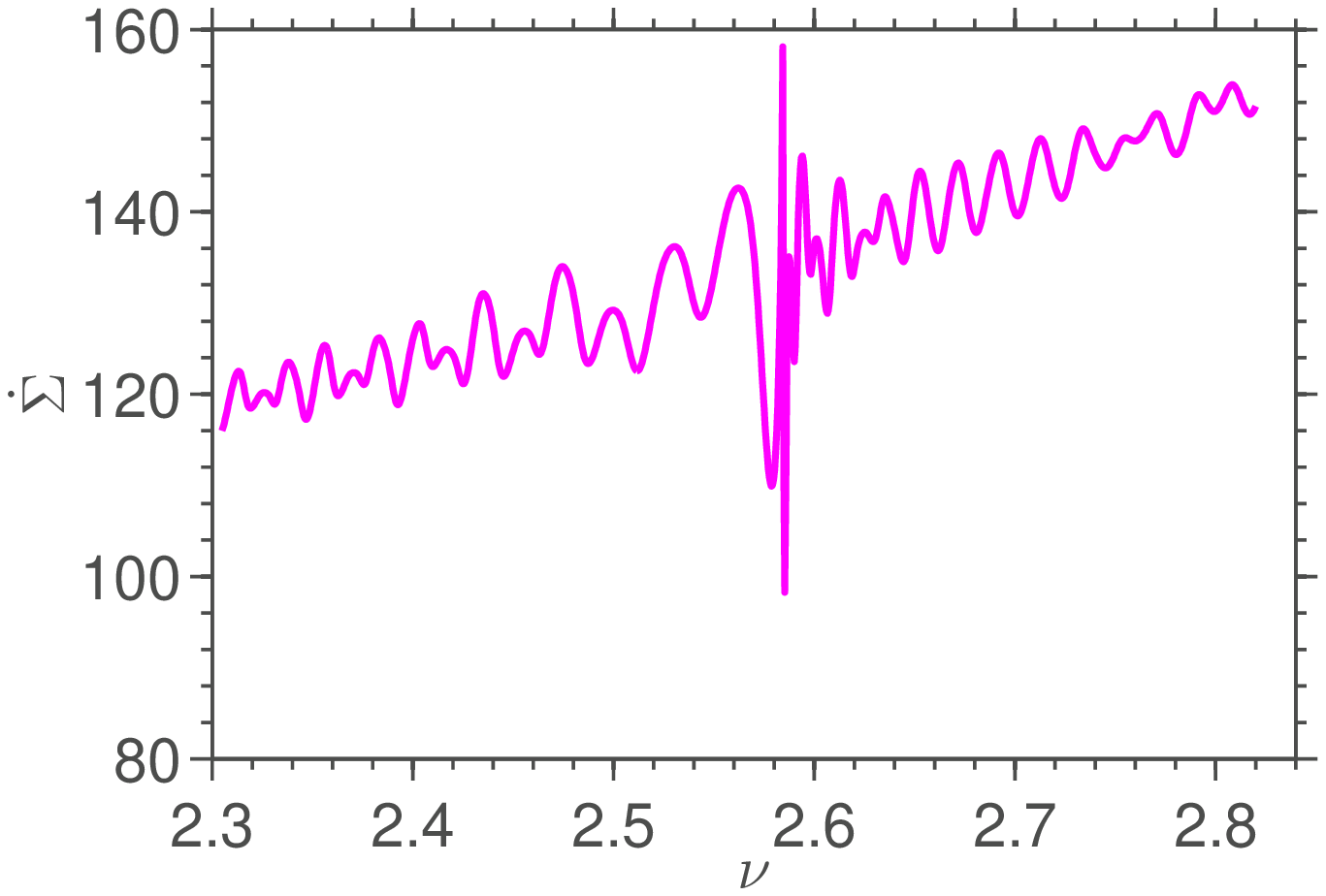}}\hfill
	\subfigure[\label{fig.totalfield}]{\includegraphics[width=0.33\linewidth]{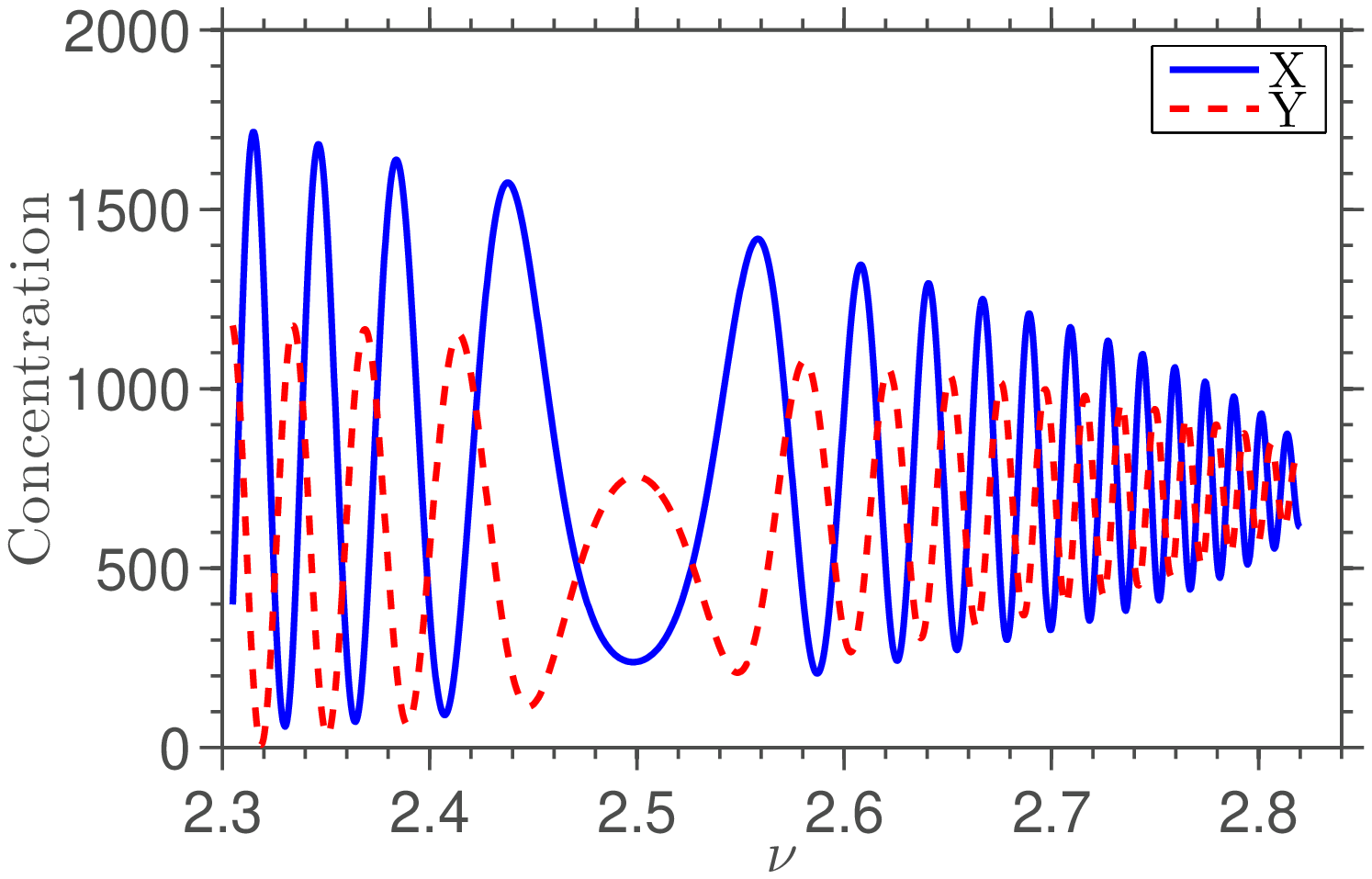}}
\hfill
	\subfigure[\label{fig.wavtotalfield}]{\includegraphics[width=0.33\linewidth]{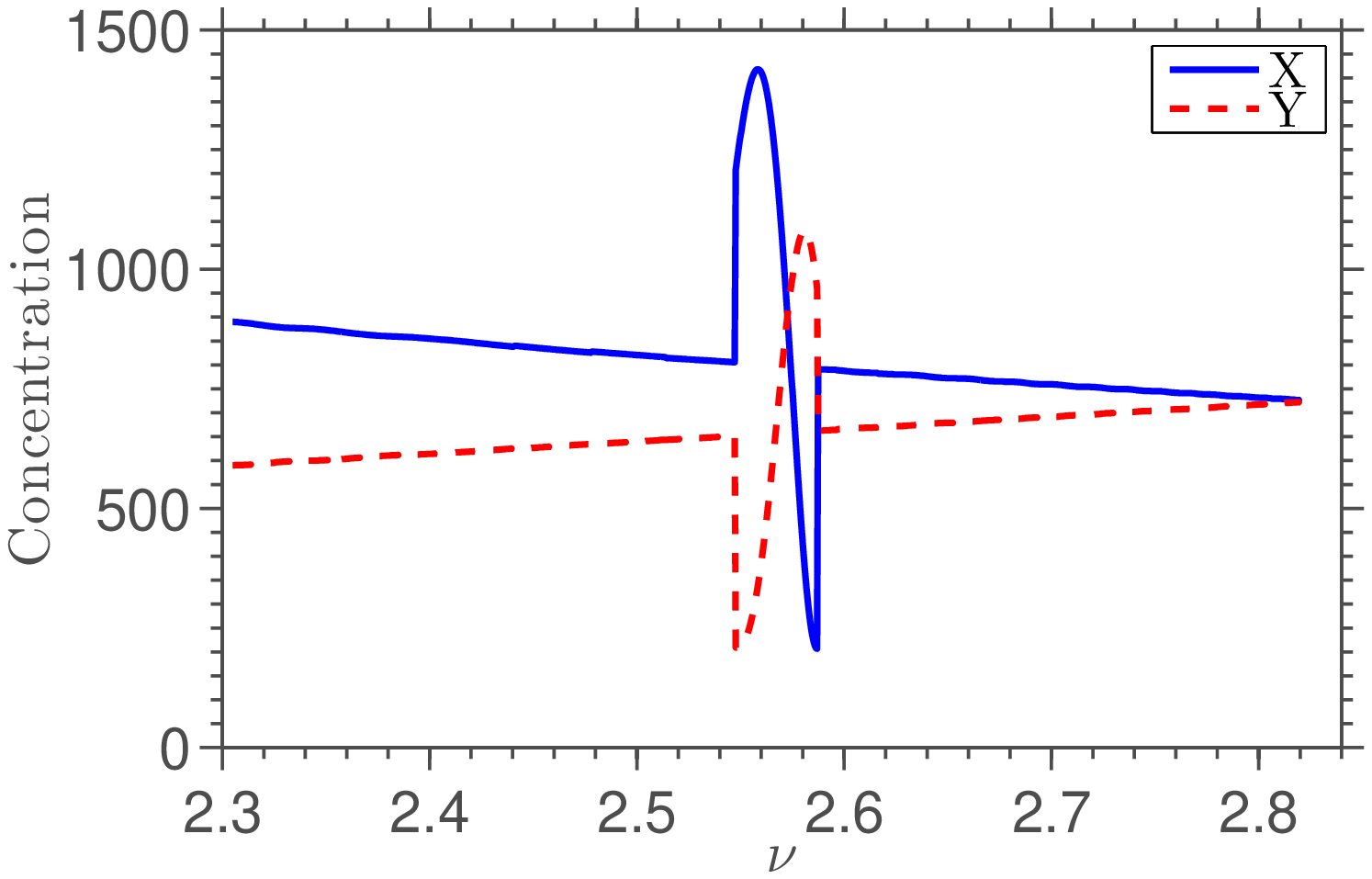}}\hfill
	\subfigure[\label{fig.wavtotalfieldin}]{\includegraphics[width=0.33\linewidth]{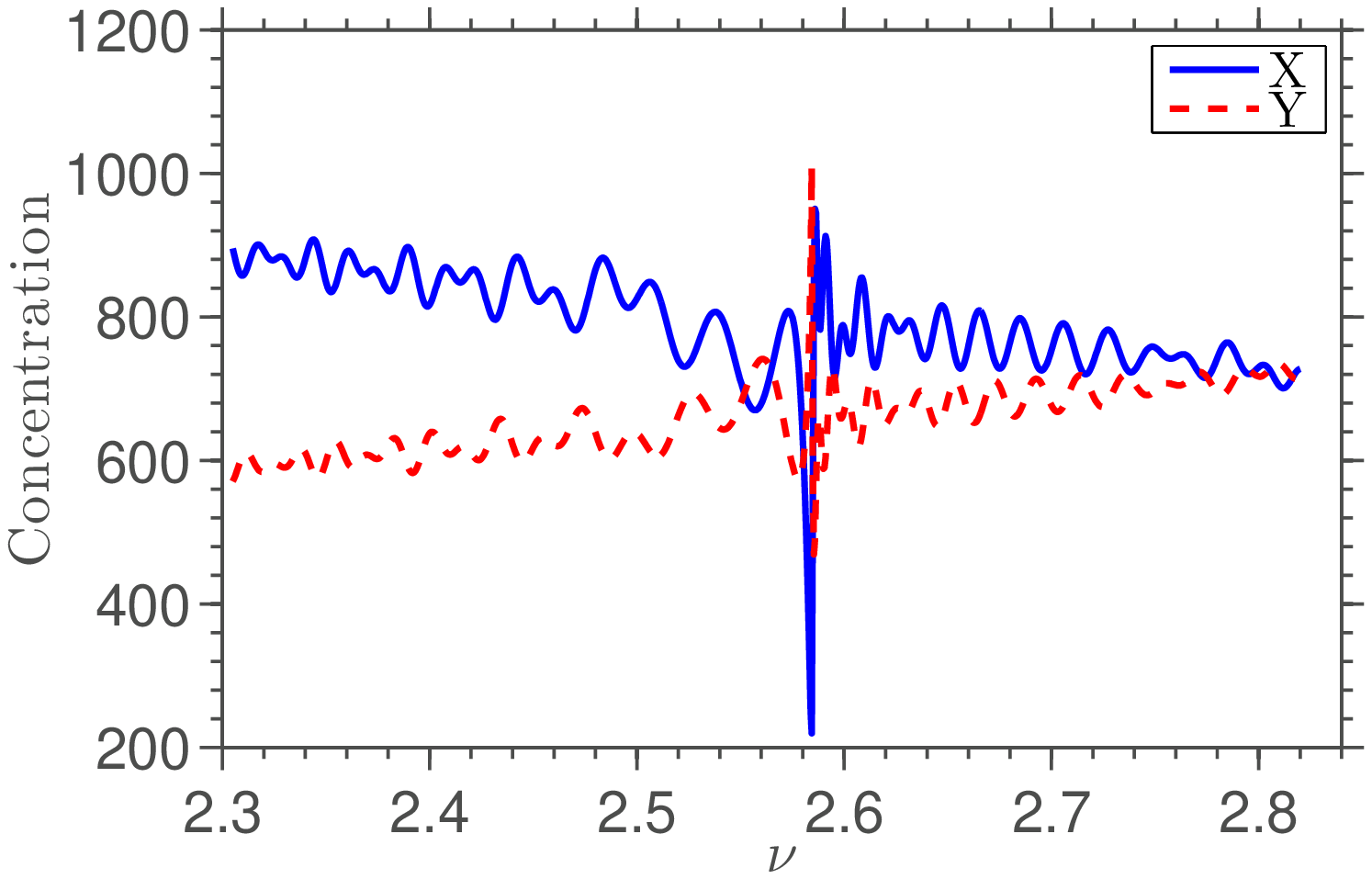}}
	\hfill
	\caption{Total entropy production of Hopf instability with respect to control parameter $\nu$ is obtained analytically for a 1D Selkov model of length $l=200$ at time, $t=400$ and absolute temperature, $T=300K$ for $\omega=2$. Total entropy production rate is comprised of  entropy production rates due to homogeneous part and reaction part. Global concentration fields of intermediate species $X$ and $Y$ as a function of $\nu$ are presented in FIG. \ref{fig.totalfield}. Total entropy production of traveling waves of the same system with discrete wavenumbers is presented in FIG. \ref{fig.wavtotalepr}. In the case of traveling waves, global concentration fields of intermediate species $X$ and $Y$ against control parameter are shown in FIG. \ref{fig.wavtotalfield}. In the limit of infinite size, the total entropy production rate and global concentration fields of intermediate species of traveling waves are presented in FIG.\ref{fig.wavtotaleprin} and \ref{fig.wavtotalfieldin}. It
  is very apparent from FIG. \ref{fig.totalepr} and \ref{fig.totalfield} and  FIG. \ref{fig.wavtotalepr} and \ref{fig.wavtotalfield} that entropy production rate is proportional to global concentration of $Y$ in terms of both magnitude and phase. For all the cases diffusion  coefficients are: $D_{11}=D_{22}=0.00051;D_{12}=-0.0002;D_{21}=0.0002$ and elementary chemical reaction  are weakly reversible i.e. $K_{-\rho}=10 ^{-4}.$} 
	\label{comp}
\end{figure*}

\begin{figure*}[htbp!]
	\centering 
	\subfigure[\label{fig.sgg}]{\includegraphics[width=0.33\textwidth]{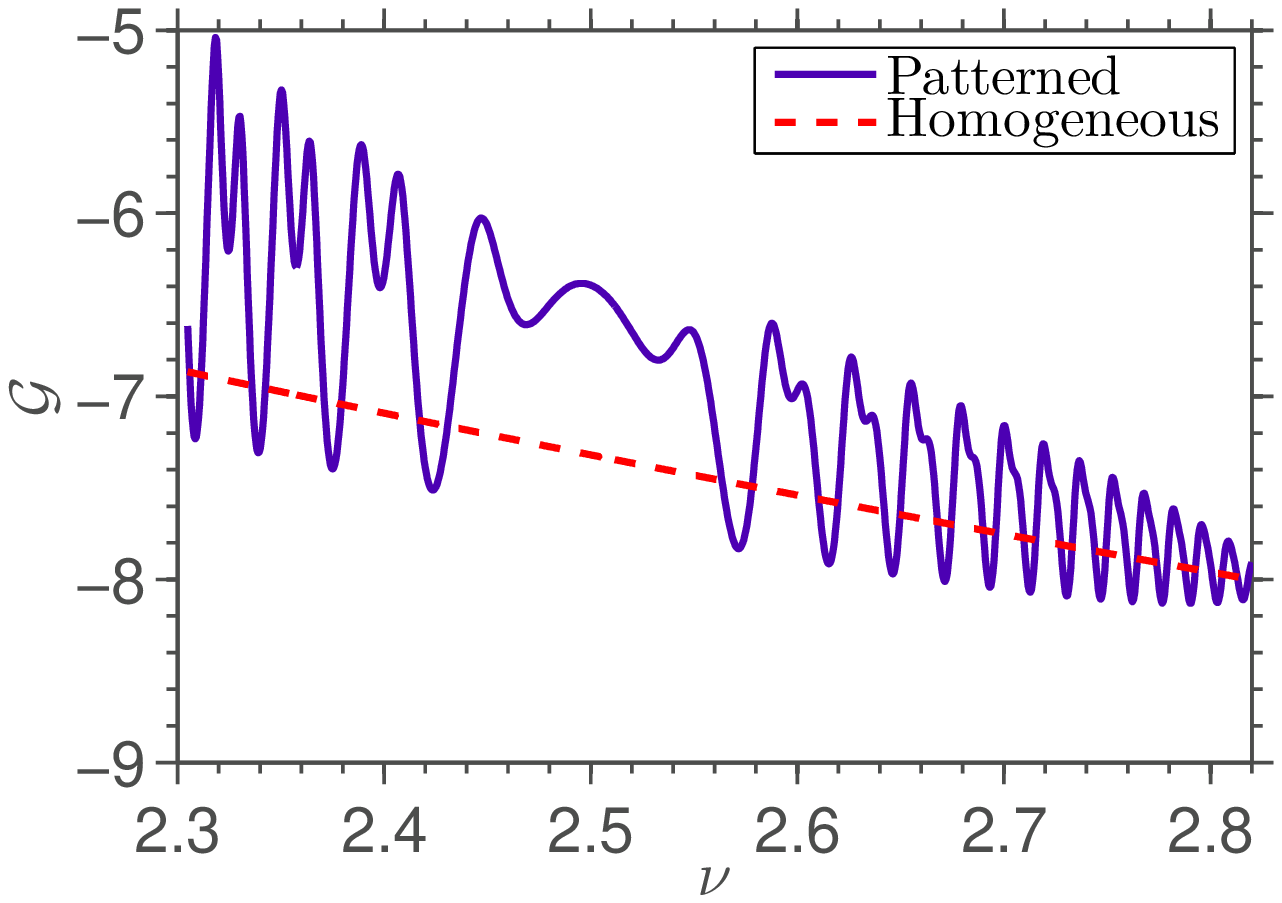}}\hfill
	\subfigure[\label{fig.wavsgg}]{\includegraphics[width=0.33\textwidth]{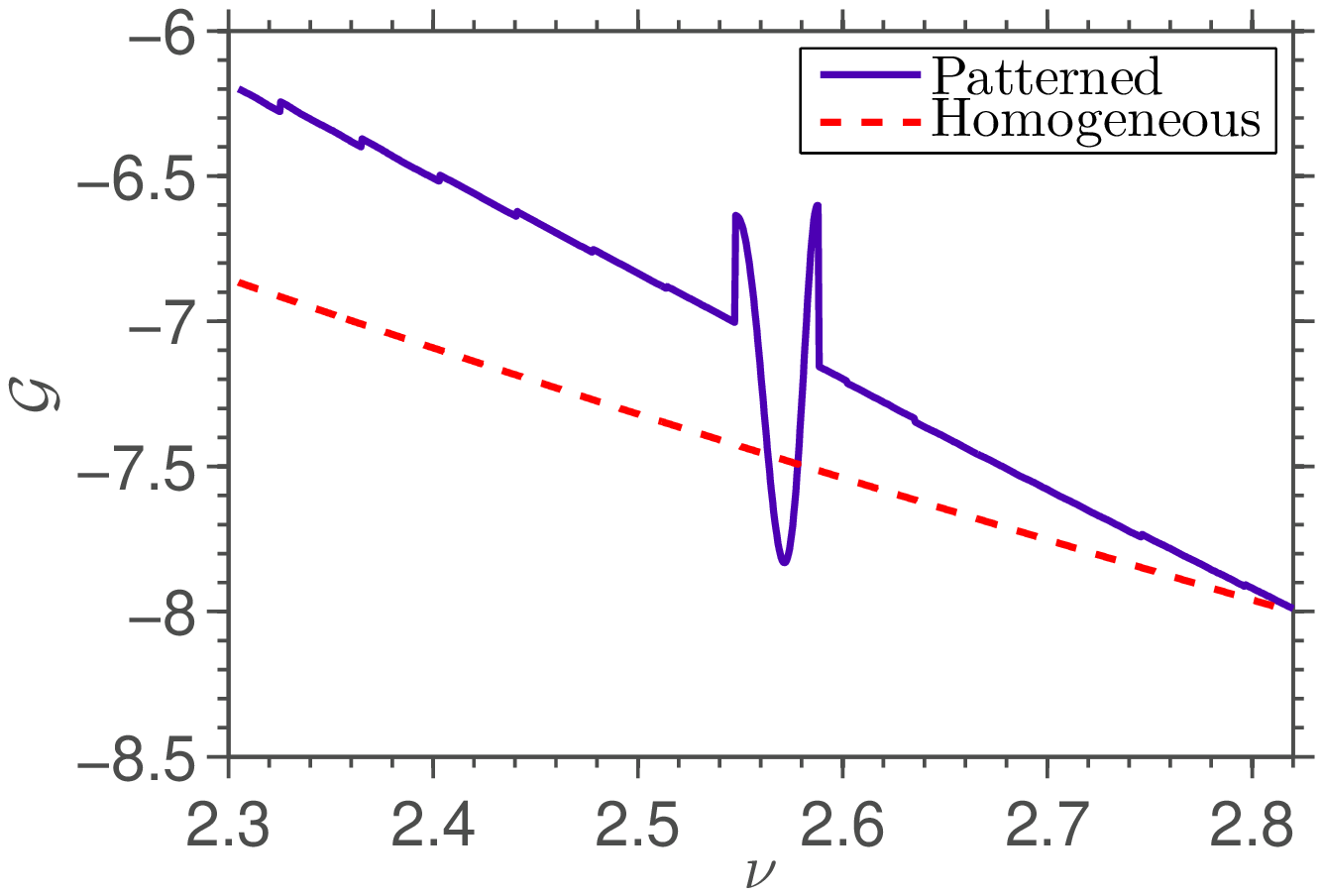}}\hfill
	\subfigure[\label{fig.wavsggin}]{\includegraphics[width=0.33\textwidth]{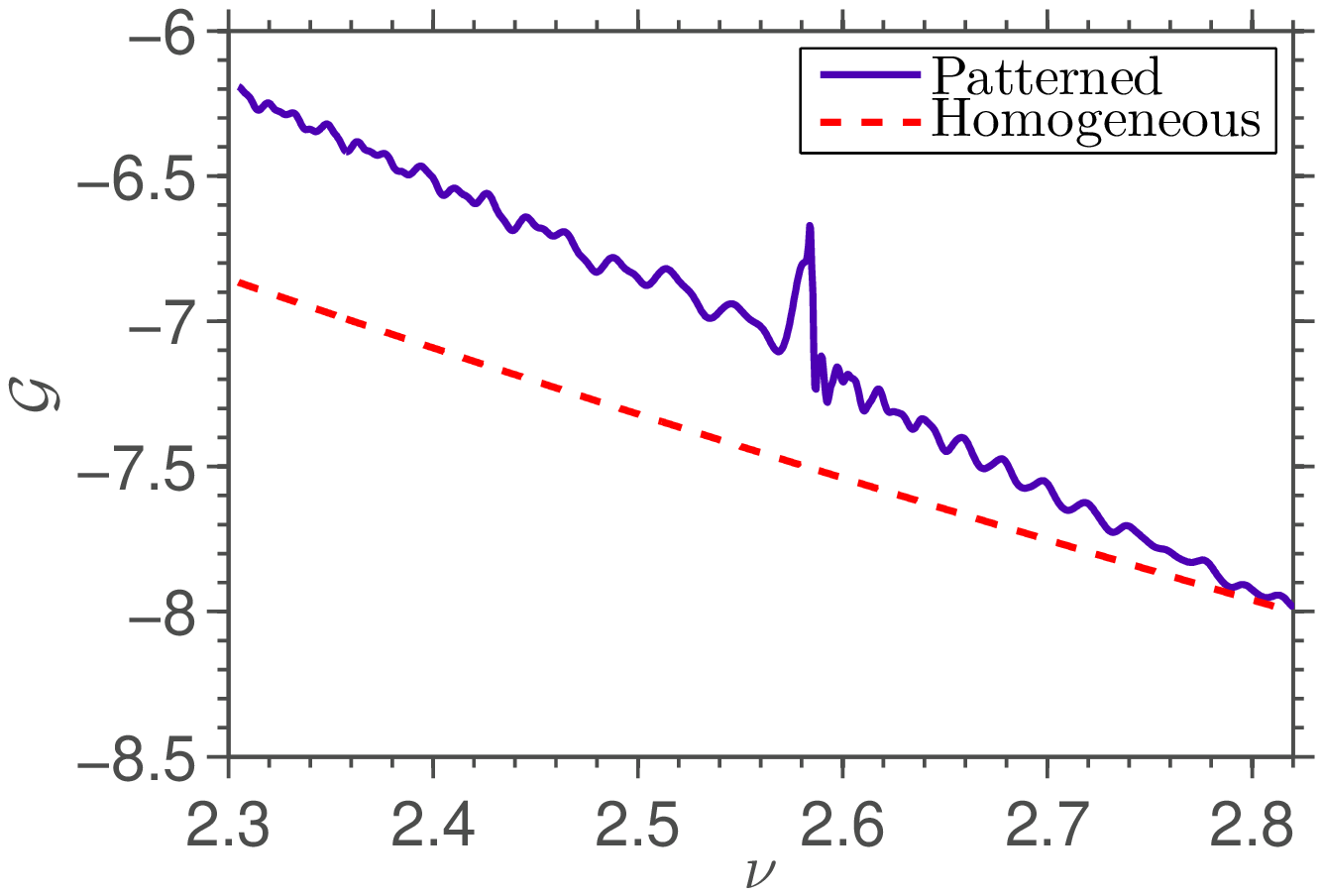}}\hfill
	\subfigure[\label{fig.slope}]{\includegraphics[width=0.33\textwidth]{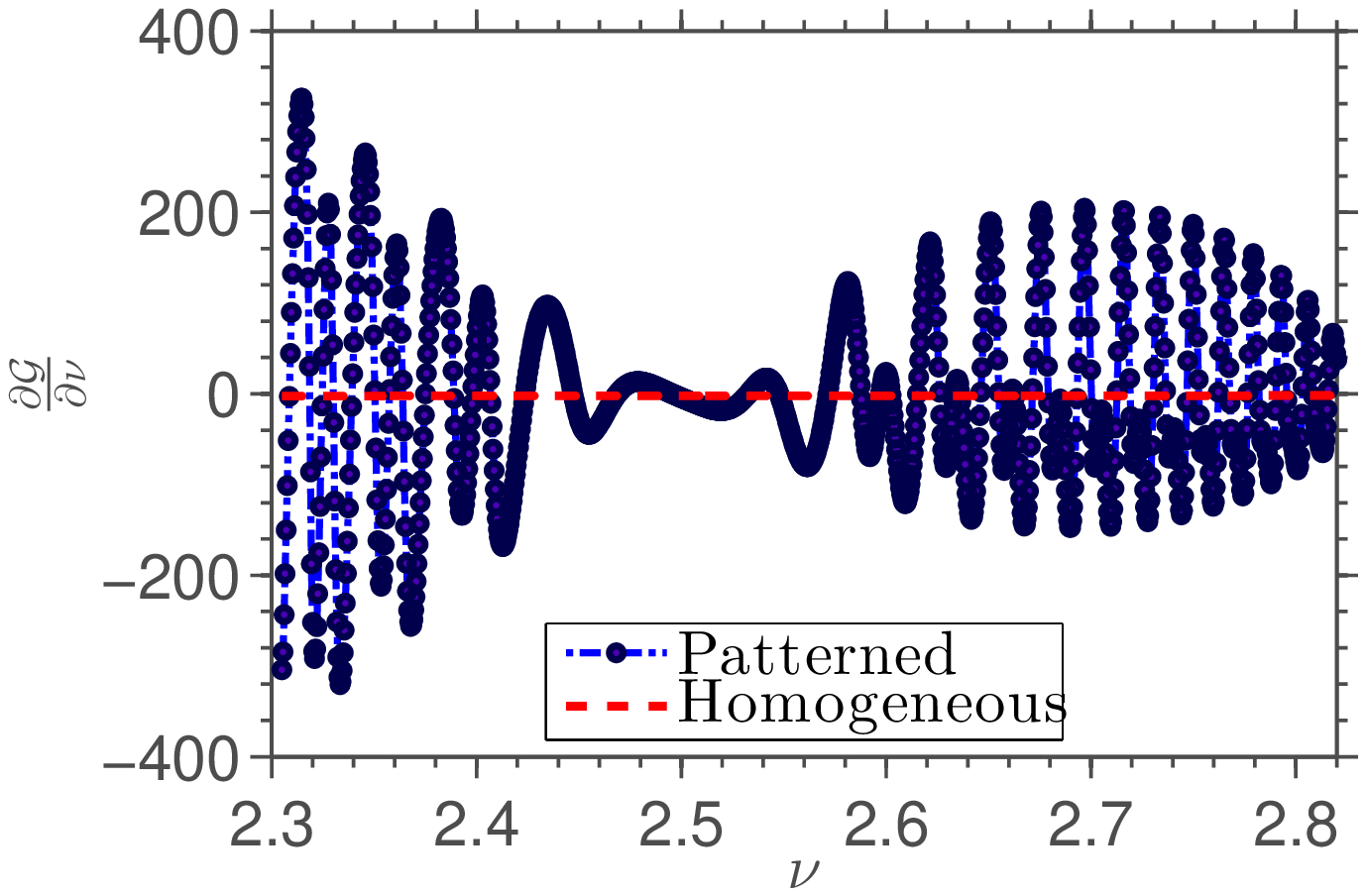}}\hfill
	\subfigure[\label{fig.wavslope}]{\includegraphics[width=0.33\textwidth]{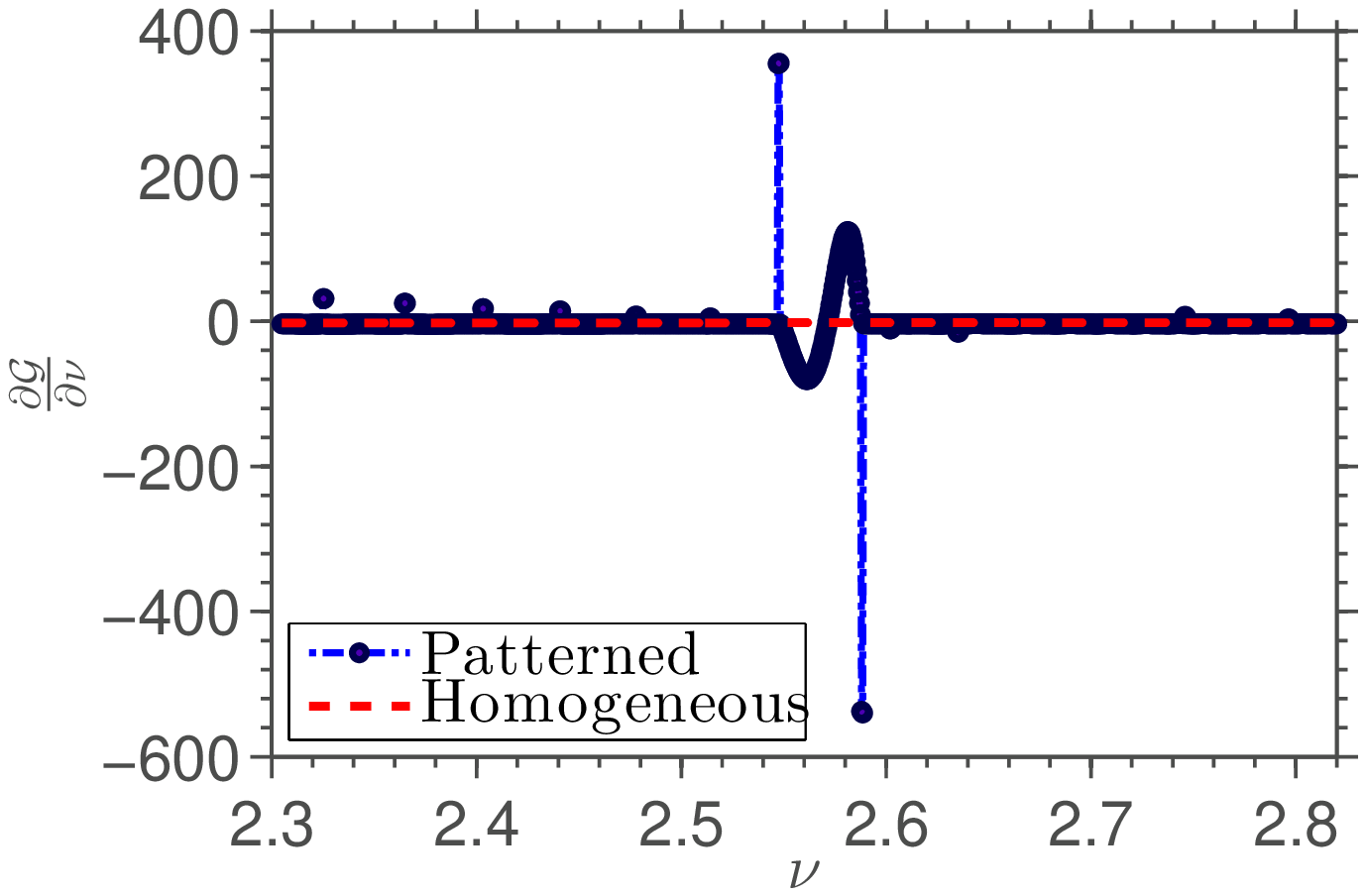}}
\hfill
	\subfigure[\label{fig.wavslopein}]{\includegraphics[width=0.33\textwidth]{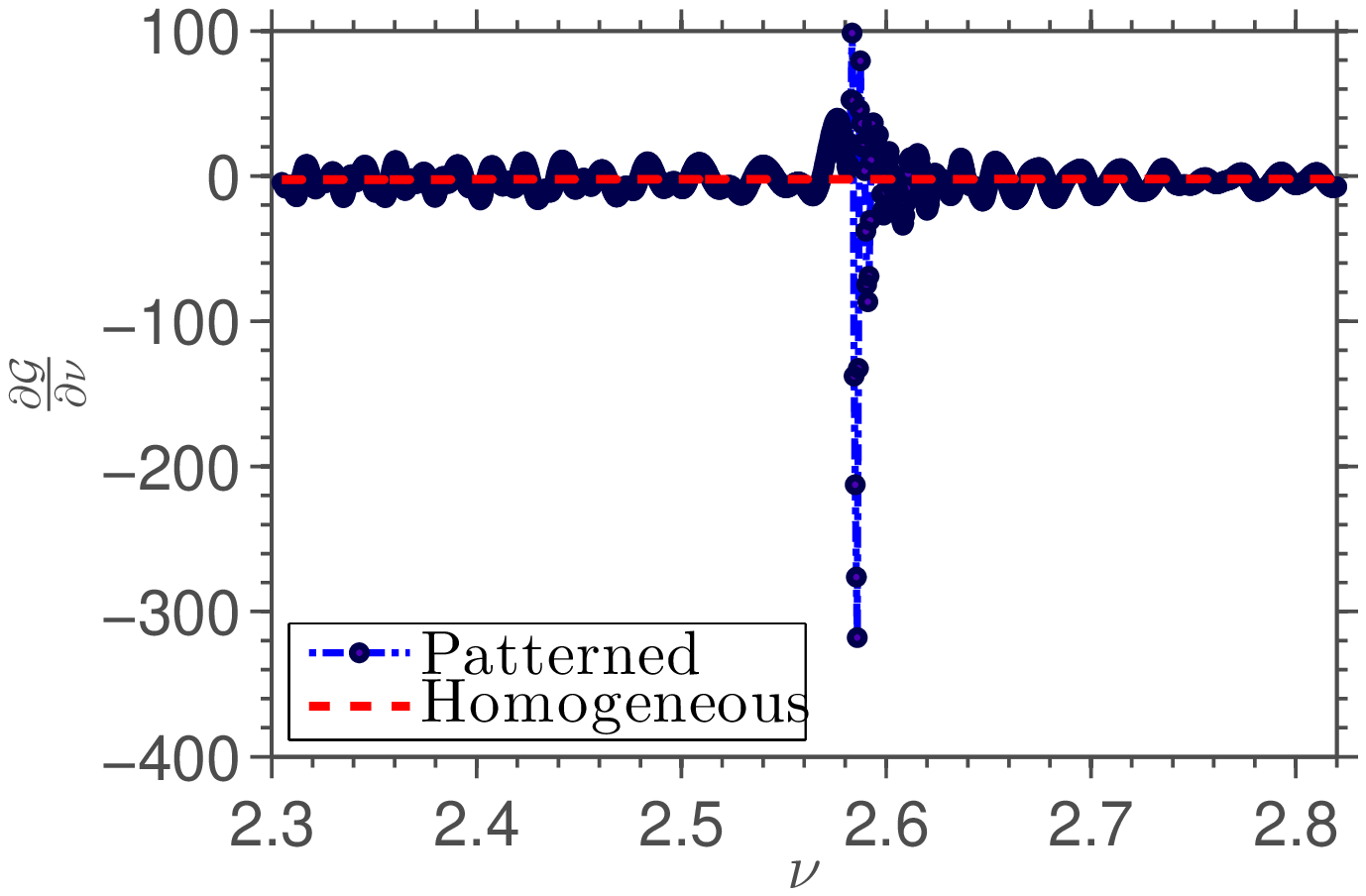}}	
\hfill
	\caption{\label{sggplusslope}The semigrand Gibbs free energy and corresponding slope of the Hopf instability are illustrated in FIG. \ref{fig.sgg} and \ref{fig.slope} respectively as function of the control parameter, $\nu$ at $t=400$ and $T=300K$ for fixed parameter $\omega=2$. The same entities for traveling waves of discrete wavenumbers are presented in FIG. \ref{fig.wavsgg} and \ref{fig.wavslope}, respectively. In the limit of infinite size, semigrand Gibbs free energy and corresponding slope profile of traveling waves are shown in FIG. \ref{fig.wavsggin} and \ref{fig.wavslopein}, respectively.  The dotted lines are for unstable homogeneous state of the system in both cases. For all the cases diffusion  coefficients are: $D_{11}=D_{22}=0.00051; D_{12}=-0.0002; D_{21}=-0.0002$.}
	\label{sgg}
\end{figure*}

The intermediate species' concentration dynamics are considered in a simple 1D glycolysis model in the regime of Hopf instability for uniform oscillation and traveling waves. Its direct correspondence in the evolution of the entities carrying the entropic and energetic descriptions of the system far from the equilibrium has been investigated systematically. All the figures here  correspond to a NESS in an one dimensional system of length, $l=200$ with absolute temperature: $T=300K$, diffusion coefficients: $D_{11}=D_{22}=0.00051; D_{12}=-0.0002; D_{21}=0.0002$,  and for weakly reversible reactions, i.e., chemical reaction rate constants $k_{-\rho}=10 ^{-4}$ unless otherwise indicated. Here, forward reaction rate constants of the eq. \eqref{revcrn} are considered as $k_{1}=k_{2}=1$ and $k_{3}=2$. As both theoretical\citep{Postnikovtemp} and experimental\citep{MAIR2005639} evidence regarding temperature dependence of glycolysis oscillation are available, one should stick to a constant temperature assumption. A constant temperature assumption implies heat diffuses much faster than the intermediate species of the reaction-diffusion model. For the parameter $\omega=2$, we would have a self-sustained oscillation for the regime below the $\nu_{cH}$. As a further expansion of the parameter space of $\nu$  results in an unphysical negative concentration of $Y$ with other parameters' values used in this analysis, we have deliberately chosen a comparatively small parametric space in the vicinity of  $\nu_{cH}$ for Hopf instability. Moreover, implementing CGLE as the backbone of our analytical investigation restricts us to near the onset of instability. We have used here discrete wavenumbers for all the analysis regarding traveling waves due to finite domain with periodic boundary condition as discussed in detail in sec. \ref{pws}.

Initially, the system is at a uniform base state set by steady-state values of two intermediate species. We have used a timestep of 0.16 and have considered 520 grid points for a system size of $l=200$. Concentration profiles of the intermediate species over the system length in FIG. \ref{xrt} for a range of control parameter values are obtained at a NESS using the analytical evolution equation of concentrations, eq. \eqref{maineq}. The total entropy production in FIG. \ref{comp} are acquired using eq. \eqref{eprr} and \eqref{eprdd} with the aid of concentrations of species at NESS. Similarly, the semigrand Gibbs profiles in FIG. \ref{sggplusslope} are obtained from eq.\eqref{smgg}.

In the lower panel of FIG. \ref{ampphasedynamics}, we have illustrated dynamics of the real part of amplitude field, $A_H$ in normalized form near the Hopf instability regime. The dynamics of $A_H$ is obtained from eq. \eqref{trialeq}(i.e., $Z$ in the equation) with the aid of eq. \eqref{steadyA} and \eqref{disperionnonlinear}. Therefore, the variation of the real part of $A_H$ reflects dynamical features of both the phase and magnitude. In the upper panel of FIG. \ref{ampphasedynamics}, variation of the phase, $\phi$  with the control parameter is demonstrated by using eq. \eqref{mainphase}. Now for the Hopf instability in FIG. \ref{fig.namp}, the selected wavenumber is simply zero, and we can observe an irregular oscillatory behavior of the real part of Hopf amplitude field, $A_H$ for a NESS at the time, $t=400$ as the control parameter, $\nu$ is varied within the Hopf instability regime. The Hopf amplitude's different magnitude concerning the control parameter implies how the corresponding limit cycle's radius gets modified. This amplitude profile helps us to understand the dynamics of local concentration in the Hopf instability parameter space at a fundamental level. FIG. \ref{fig.hopfphase} shows the system's phase change for the Hopf instability as a function of the control parameter. The comparison between FIG. \ref{fig.namp} and FIG. \ref{fig.hopfphase} suggests that corresponding to a `double-well' shaped region of low amplitude  at $\nu=2.5$ in FIG. \ref{fig.namp}, the phase of the system passes through a local minimum. Therefore we can state that the appearance of the `double-well' shaped low-amplitude region in the normalized amplitude profile is due to phase minima at that point in the case of Hopf instability. Now for traveling waves, initial changes in the normalized profile are mainly due to quantization of wavenumber(see FIG. \ref{fig.wavenumber}). One can notice no notable change in the normalized real part of the amplitude field of FIG. \ref{fig.wavnamp} while phase passes through the minimum in FIG. \ref{fig.wavephase}. However, when the spatial distribution of the system phase ceases to a single point in FIG. \ref{fig.wavephase}, a spatially homogeneous part in the normalized amplitude in FIG. \ref{fig.wavnamp} appears near $\nu=2.6$. In FIG. \ref{fig.namp} and \ref{fig.wavnamp}, we can also observe that magnitude exhibits a gradual increase relatively far from the onset of oscillatory instability point for the variation of the control parameter. The phase and amplitude change for traveling waves in the limit of infinite size, i.e., for continuous wavenumber band are illustrated in FIG. \ref{fig.wavephasein} and \ref{fig.wavnampin}, respectively. Both phase and amplitude of traveling waves with continuous wavenumbers exhibit more smooth transitions than their discrete wavenumber counterpart. 

The 3d concentration field and a corresponding image of the intermediate species, $X$ as obtained from the analytical eq. \eqref{maineq} are presented in FIG. \ref{xrt}. One can notice an extra-wide region of low concentration between the control parameter value $\nu=2.4$ and $\nu=2.6$ in FIG. \ref{fig.ximage}. The extra-wide region of low concentration has a one-to-one correspondence with the  `double-well' shaped low-amplitude. We also discussed that the low-amplitude profile is connected with the phase slope's sign change from negative to positive. To understand the wide low concentration region area in terms of the amplitude equation's coefficients, we need to consider FIG. \ref{fig.xline}. The red dotted line of the FIG. \ref{fig.xline} corresponds to the variation $\alpha+\beta$ of the CGLE with control parameter $\nu$  and the dotted black line refers to $\alpha+\beta=0$ or $1-\alpha \beta=0$ condition. As discussed in the sec. \ref{sec:hae}, whenever the $\alpha +\beta$ values crosses the $\alpha+\beta=0$ line the phase of the spiral would be reversed. However, the $\alpha+\beta$ profile always remains below the zero line in FIG. \ref{fig.xline}. Here, the solid blue line in the FIG. \ref{fig.xline} represents the $1-\alpha \beta$ condition. The $1-\alpha\beta$ profile crosses the zero line near $\nu=2.6$ and signals a transition from the uniform oscillation to BF instability with the control parameter variation. Now, one can notice that the amplitude in FIG. \ref{fig.namp} demonstrates an abrupt low amplitude profile before the point of onset of BF instability, and this low amplitude profile dictates the extra-wide low concentration region of the concentration field shown in FIG. \ref{fig.ximage}. As the control parameter value is increased beyond the onset of instability point, there is a damped oscillation type of behavior. Moreover, corresponding to the extra-wide low concentration regime in the $X$ concentration of the Hopf instability, we can observe a clear turn back before moving towards the center in the illustration, similar to a phase portrait in FIG. \ref{fig.limit}. Therefore, the direction change in the phase portrait of the Hopf instability is a consequence of a phase slope sign at the fundamental level.  Lavrova et al.\citep{LAVROVA2009127} found a pulsating regime in temporal dynamics of $X$ concentration within the uniform oscillatory regime corresponding to phase reversal with time in a previous study of glycolytic waves propagation with inhomogeneous substrate influx. 

Instead of the zero wavenumber of uniform oscillation, when we consider a small finite wavenumber near the $q_{cH}=0$, we would obtain a traveling wave type concentration profile within the oscillatory regime of the system. In FIG. \ref{fig.wavenumber}, wavenumber near the critical wavenumber of the Hopf instability is shown as a function of the externally controlled parameter, $\nu$.  Discrete wavenumbers for different control parameter values that have been used as allowed wavenumbers for the traveling wave can be seen from the bold line the FIG. \ref{fig.wavenumber}. These discrete values of wavenumber are allowed in the finite domain with periodic boundary conditions and are obtained with the aid of analytically derived continuous wavenumber expression eq. \eqref{selected_wavenumber} and \eqref{crtick}. The wavenumber decreases and approaches the critical value of wavenumber for  Hopf instability as we change the control parameter value towards the onset of the BF instability point while all other parameters remain constant. Then we have again considered discrete and finite wavenumbers within the BF instability region. In the same figure, a dashed line shows the corresponding continuous wavenumber profile, which can be selected as valid wavenumbers in the limit of infinite system size.  The 3d concentration field of $X$  for the traveling waves with discrete wavenumbers is shown in FIG. \ref{fig.wavx} and the corresponding image of the concentration field are presented in FIG. \ref{fig.wavximage}. 
In the stable oscillatory regime of the control parameter, traveling waves are demonstrated for discrete wavenumbers within linearly stable wavenumber limit in FIG. \ref{fig.wavximage}. Even when the control parameter goes beyond the onset of BF instability, as seen in FIG. \ref{fig.xline}, our selection of nonzero wavenumber generates a modulated pattern due to the convective nature of BF instability as seen in  FIG. \ref{fig.wavximage}. The disconnectedness in the traveling wave pattern is due to the different discrete wavenumber values corresponding to the control parameter values. Modification in traveling waves pattern around the BF instability due to the control parameter variation is dictated by the change in amplitude dynamics while passing through the onset of the BF instability point. For traveling waves, by selecting the zero wavenumber around the onset of BF instability as seen from FIG. \ref{fig.wavenumber},  amplitude dynamics have a spatially homogeneous part in FIG. \ref{ampphasedynamics}. Although the wavenumber of the traveling waves at the onset of BF instability is equal to the Hopf instability wavenumber, we need to treat them as two different dynamic features. For the infinite size limit, wavenumbers of traveling waves can have continuous values, as shown in FIG. \ref{fig.wavenumber}. For continuous wavenumber, the concentration pattern of traveling waves is different from its finite domain counterpart. The 3d concentration field of $X$ and the corresponding image of the concentration field for the infinite size limit are shown in FIG. \ref{fig.wavxin} and \ref{fig.wavximagein}, respectively. The difference in the concentration pattern of intermediate species for the discrete wavenumbers and the continuous wavenumbers is predictable given their different amplitude dynamics. Therefore, incorporating the idea of different instability conditions directly related to amplitude equation coefficients, $\alpha$ and $\beta$ with the amplitude dynamics concerning $\nu$ in the FIG. \ref{ampphasedynamics}, we can understand and predict the profiles in the FIG. \ref{xrt} better. The $\alpha$ parameter playing an important role in setting instability conditions in the amplitude framework contains both the self and cross-diffusion coefficients. Thus, the cross diffusion can alone shift the parametric regimes shown in FIG. \ref{fig.xline} and thus modify the temporal concentration pattern to a significant extent, especially in the case of equal self-diffusion coefficients.

The entropy production rate owing to its origin to reaction and diffusion in the system is obtained separately using eq. \eqref{eprr} and \eqref{eprdd}, respectively in the presence of cross-diffusion.  Even in the traveling wave case where the wavenumber selection directly depends on the parameter containing diffusion coefficients,  there is no significant contribution from the diffusion part to the entropy production. Therefore total entropy production is basically due to the sum of the initial homogeneous part and reaction dynamics of the global system. 

We have investigated the total EPR response due to the variation in the control parameter, $\nu$ keeping another parameter, $\omega$ constant in both Hopf and traveling wave cases. For Hopf instability, a nonzero total EPR shows oscillatory response, as shown in FIG. \ref{fig.totalepr}. Comparing profiles of the global concentration of $X$ and  $Y$ in FIG. \ref{fig.totalfield} and corresponding total EPR in FIG. \ref{fig.totalepr} one observes that the total EPR  is quantitatively proportional to the global concentration of $Y$. Moreover, they have qualitatively similar dynamics. In other words, the total EPR  reflects the global dynamics of  $Y$ concentration. This similarity implies that we can exploit the total EPR of a dissipative system as a quantitative and qualitative measure of the system's temporal pattern.  Traveling waves' total EPR in FIG. \ref{fig.wavtotalepr}, exhibits a pulse-type response around the onset of BF instability point. The total EPR of traveling waves otherwise shows a clear upward trend against the control parameter as seen in FIG. \ref{fig.wavtotalepr}. The oscillatory nature of traveling waves is not prominent for discrete wavenumber cases. The total EPR of traveling waves in the limit of infinite system size exhibits the oscillatory nature of traveling waves over the whole range of $\nu$ and only suffers an abrupt sharp change around the onset point of BF instability. Similar to the Hopf instability, the total EPR  of the traveling wave is analogous to the global dynamics of  $Y$ concentration in FIG. \ref{fig.wavtotalfield} and \ref{fig.wavtotalfieldin}. Thus, a dissipative system's total EPR can capture the temporal and spatial inhomogeneities both quantitatively and qualitatively, irrespective of its size.   

FIG. \ref{fig.sgg} illustrates the semigrand Gibbs free energy change as a function of the control parameter $\nu$  for Hopf instability. As suggested by the FIG. \ref{fig.sgg},  semigrand Gibbs free oscillates around its unstable homogeneous counterpart. The semigrand Gibbs free energy has a $(2:1)$ periodic oscillation feature. The extra-wide low concentration regime of $X$ concentration field can also be identified around $\nu=2.5$ as a comparatively slow change in the semigrand Gibbs free energy profile. The plot of slopes for the same thermodynamic entity is shown in the FIG. \ref{fig.slope} and as expected it confirms the slow variation of semigrand Gibbs free energy around $\nu=2.5$. For traveling waves in a finite domain with periodic boundary conditions, the semigrand Gibbs free energy against $\nu$ is demonstrated in FIG. \ref{fig.wavsgg} and corresponding slopes are shown in FIG. \ref{fig.wavslope}. Like the total entropy production rate, the oscillatory behavior of traveling waves with discrete wavenumber is not clear enough in the energetic entity. However, the change in the semigrand Gibbs free energy due to the spatial pattern generation is visible as it separates the semigrand Gibbs free energy profile of the traveling wave from the unstable homogeneous counterpart. The semigrand Gibbs free energy for the traveling wave is greater than the system's unstable homogeneous counterpart except around the onset of the BF instability point. The increase in semigrand Gibbs free energy for the spatial pattern is due to the work needed to vary the wavenumber of the traveling wave.  Around the onset of BF instability point, the wavenumber of a traveling wave for finite system size is equal to the critical wavenumber to Hopf instability, and the semigrand Gibbs free energy decreases from its value for the unstable homogeneous state and passes through a minimum. Unlike the discrete wavenumber case, the semigrand Gibbs free energy in the infinite size limit has a prominent oscillatory behavior and exhibits a clear maximum at the onset of BF instability.

Slopes of semigrand Gibbs free energy in the Hopf instability and traveling waves are of the same order in FIG. \ref{sggplusslope}. Akin to the total entropy production rate, a pulse-like behavior appears around the onset of BF instability in the slope of semigrand Gibbs free energy for the finite system size in FIG. \ref{fig.wavslope}. A close similarity in the profiles of entropy production rate and slope of semigrand Gibbs free energy is also observed for the infinite size limit, as shown in FIG. \ref{fig.wavslopein}. These similarities suggest that the slope of semigrand Gibbs free energy is proportional to the total entropy production rate of the system. 

\section{\label{conclu}Conclusion}

Capturing the uniform oscillation and traveling wave dynamics of the system are implemented here by CGLE based description to a more general reaction-diffusion system in the presence of cross-diffusion. Then opting for a rigorous nonequilibrium framework for entropic and energetic characterization of the temporal and spatial dynamics of the system, we have provided a general recipe for relating any dynamic signature with nonequilibrium thermodynamic entities explicitly. Besides the uniform oscillation and traveling waves,  this  analytical study applies to any pattern or overlapping of different patterns\citep{DeWitA1996SpatiotemporalPoint., Just2001Spatiotemporal, Yang2003OscillatoryLayers} within a more general environment or the spatiotemporal dynamics owing to BF instability.

As the amplitude equation explicitly contains all the diffusion matrix elements, diffusion coefficients affect the amplitude and phase of the system through the form of a CGLE solution. Besides, the wavenumber also has implicit cross diffusion dependence, which is again reflected by the coefficients of the amplitude equation. Here, the challenging task of wavenumber selection in the nonequilibrium system has been handled by obtaining a boundary value of linearly stable wavenumber through the perturbation method of testing the stability of plane waves and then modifying it for finite domain case by considering admissible discrete wavenumber values.   

We have restricted ourselves to the global thermodynamic description of Hopf instability and traveling waves in this report. As all conservation laws of the closed system are broken here by chemostating in the corresponding open system, the semigrand Gibbs free energy is equivalent to the system's energetic entity at the local level \citep{chemicalwaveespasito}. 
Our previous study\citep{pkgg} found proportionality of total EPR  with global concentration profile in Turing-Hopf overlapping regime.  Here, we have obtained that EPR dynamics is analogous to the global concentration dynamics both qualitatively and quantitatively for uniform oscillation and traveling waves. We have also acquired a pulse-like shape in the total EPR profile at the onset of BF instability for the finite wavenumbers. However, in the limit of infinite system size with continuous wavenumber, the total EPR profile demonstrates a sharp change at the onset of BF instability. 

We have found that the energetics of the Hopf instability over the whole control parameter range is more complicated than the traveling wave. Here the appearance of a 
$(2:1)$ periodic oscillation in semigrand Gibbs free energy of the Hopf instability is an example of nonlinear resonance. Surprisingly, we have obtained different natures of the semigrand Gibbs free energy around the BF onset associated with discrete and continuous wavenumber consideration. This contradiction indicates the significance of considering the finite boundary effect for a thorough investigation of the traveling wave in a real situation.

Here, we have considered sufficiently small but equal self-diffusion coefficients\citep{Lavrova2009PhaseInflux, LAVROVA2009127, Verisokinpre} in the presence of cross-diffusion coefficient to generate complex oscillation patterns.    
Thermodynamic description of Hopf instability and the traveling wave can be extended to control the collective dynamics of biological, physical, or chemical oscillators from a new perspective\citep{strogatz}. It is also possible to extend this analysis to study the spiral waves\citep{winfree1972spiral} and their phase reversal scenario\citep{Vanag835, Gong, chao}.  We believe this CGLE based framework can be applied even in the super or sub diffusive regime as well as in the presence of concentration dependent diffusion. However, it is noteworthy that CGLE may be questionable\citep{shao} to capture the anti-wave to wave transition as the transition can happen away from the Hopf onset point. In a similar context, the thermodynamic framework for non-elementary chemical reaction network\citep{Avanzini_2020} is also relevant.

\bibliography{selkovreport}
\end{document}